# Atomistic insight into defect-mediated boron redistribution and high-temperature strengthening in TiAl alloys

Xiang Meng[1], Shuhei Shinzato[2], Zhiqiang Zhao[2*], Linfeng Yu[1], Jian Guo[1], Liangli Chu[1], Kepeng Ouyang[1], Shigenobu Ogata[2*] and Zhuhua Zhang[1*]

[1] *State Key Laboratory of Mechanics and Control for Mechanical Structures, College of Aerospace Engineering, Nanjing University of Aeronautics and Astronautics, Nanjing 210016, P. R. China*

[2] *Department of Mechanical Science and Bioengineering, Graduate School of Engineering Science, The University of Osaka, Osaka 560-8531, Japan*

[*]Corresponding authors: zhiqiang.zhao@tsme.me.es.osaka-u.ac.jp (Zhiqiang Zhao); ogata.shigenobu.es@osaka-u.ac.jp (Shigenobu Ogata); chuwazhang@nuaa.edu.cn (Zhuhua Zhang)

**Abstract:** Boron (B) microalloying can effectively refine microstructures of as-cast TiAl alloys and regulate their mechanical properties. Yet, how dilute B exists in TiAl and interacts with defects and how these interactions govern high-temperature deformation remain elusive. Here, we systematically elucidate defect-mediated B redistribution in TiAl alloys and its consequences for their deformation using hybrid grand-canonical Monte Carlo/molecular dynamics simulations augmented with a newly developed high-fidelity machine-learning potential. As a typical product of B microalloying, borides are markedly stiffer than the TiAl matrix and act as brittle reinforcing phases, while borides/matrix interfacial adhesion depends on surface termination and stacking sequence. In the bulk, interstitial B tends to be trapped at Ti-rich octahedral sites by high migration barriers but can kinetically evolve into substitutional dopants upon approaching Al vacancies. While increased tensile stress and vacancy concentration further promote B enrichment in the bulk, B is enriched at semi-coherent $\gamma/\alpha_2$ boundaries, and partitions to $\alpha_2$. In polycrystalline TiAl, B accumulates primarily in disordered regions, raising the average flow stress by 9.4% at 1 at.% B and 900 K. This strengthening is accompanied by reduced shear strain accumulation. Simulated microstructural evolution and stacking-fault energetics indicate that B impedes dislocation glide via local pinning and suppresses deformation twinning, thereby increasing resistance to plastic deformation. By linking solute redistribution to defect interactions and deformation, these findings establish an atomistic framework for understanding and designing high-performance alloys based on interstitial solutes.

**Keywords: TiAl alloys, boron alloying, machine-learning potential, interfacial segregation, high-temperature strengthening**

## 1. Introduction

$\gamma$-TiAl based alloys are attractive lightweight structural materials for high-temperature turbine components (1-3), owing to their low density and unique combination of high specific strength and excellent resistance to creep and oxidation (4, 5). Their broader deployment, however, remains constrained by the strong dependence of plasticity and damage tolerance on microstructure heterogeneity. Conventional cast TiAl alloys, particularly those undergoing peritectic solidification, commonly develop coarse columnar grains, chemical segregation and nonuniform lamellar colonies. These microstructural features introduce pronounced colony-scale anisotropy and local compositional inhomogeneity, hindering coordinated plastic deformation and promoting strain localization. The resulting deformation incompatibility accelerates damage accumulation and crack initiation, reducing both service strength and room-temperature ductility (6). At elevated temperatures, interface migration and structural evolution further govern deformation and failure, making the local interfacial environment a key factor controlling mechanical performance. Improving TiAl alloys therefore requires not only refinement of the cast microstructure, but also atomic-scale control of the interfacial chemistry and defect activity that govern plasticity and failure.

Minor boron (B) alloying provides an effective approach for refining as-cast TiAl microstructures without costly post-casting heat treatments (7). For example, B addition promotes grain refinement in Ti45Al2Mn2Nb alloys under rapid solidification conditions (8). This refinement arises from the low solubility of B in TiAl phases, which drives B partitioning into the liquid during solidification and suppresses the continuous growth of coarse columnar grains. Meanwhile, Ti-B compounds such as TiB and $TiB_2$ (9), when formed at appropriate solidification stages, can further promote heterogeneous nucleation and contribute to grain and colony refinement. The extent of this refinement also depends on B content. For example, in Ti47Al2Cr2Nb alloys, significant grain refinement was observed only when the B content exceeded approximately 0.5 at.% (10). These observations establish the dominant role of B in refinement, whose effectiveness is governed by solute rejection and boride-assisted nucleation. Yet the distribution of residual B outside borides, and its mechanical role in segregation, trapping and defect-mediated deformation, remain poorly understood.

Beyond grain refinement and precipitation strengthening, B may directly modify mechanical behavior by regulating interfaces and defects at the atomic scale. Experiments on steels (11), Ni-based intermetallic alloys (12), and complex concentrated alloys (13, 14) have shown that B tends to segregate to grain boundaries, thereby enhancing boundary cohesion and modifying crack behavior. Such effects are particularly relevant to TiAl alloys, where plasticity and failure are strongly governed by dense $\gamma/\gamma$ and $\gamma/\alpha_2$ lamellar interfaces. However, it remains unclear how B interacts with diverse interfacial structures and deformation-induced defect environments in TiAl alloys. Direct experimental characterization is challenging because B is highly dilute and its local environment can evolve during deformation through diffusion, vacancy trapping and interfacial redistribution. First-principles calculations have revealed the local chemical preference of B for Ti-rich interstitial sites and $\gamma/\alpha_2$ interfaces (15, 16), but cannot dynamically track the large-scale defect evolution under applied loading. Molecular dynamics (MD) simulations provide the required computational efficiency, yet their application to Ti-Al-B is limited by the absence of a dedicated empirical or machine-learned potential capable of faithfully describing complex Ti-Al-B bonding environments and B-matrix interactions during deformation.

Here, we address these gaps by developing a high-fidelity machine-learning potential for the ternary Ti-Al-B system and combining it with hybrid grand-canonical Monte Carlo/molecular dynamics (GCMC/MD) simulations to resolve the defect-mediated redistribution and mechanical role of dilute B in TiAl alloys. The potential reproduces key bulk, surface, interfacial and defect properties, including elastic constants, stacking fault energetics, B migration barriers, and B-vacancy interactions, providing a consistent description across atomic and microstructural scales. Using this potential, we investigate B incorporation at matrix interstitial sites, trapping by vacancies and segregation at interfaces at 900 K. Interfacial cleavage, stacking-fault and polycrystalline tensile simulations further connect these interactions to interfacial cohesion, deformation twinning and high-temperature mechanical response. Our findings identify Ti-rich octahedral sites and Al vacancies as important trapping environments for residual B and suggest that its defect-mediated redistribution contributes to enhanced high-temperature deformation resistance in TiAl alloys.

## 2. Models and methods

### 2.1 DFT calculations and dataset construction

All reference data for potential development were generated with the VASP package (17) using the projector augmented-wave (PAW) method and the Perdew-Burke-Ernzerhof (PBE) exchange-correlation functional within the generalized gradient approximation (18). A plane-wave cutoff of 600 eV was used, with $\Gamma$-centered $k$-point meshes of 0.15 $Å^{-1}$ grid spacing and first-order Methfessel-Paxton smearing of 0.1 eV width. The electronic self-consistency criterion was $10^{-5}$ eV, and the structures were relaxed until the residual force on every atom fell below 0.01 eV/Å. All calculations were non-spin-polarized. The resulting energies, atomic forces and virials are the reference targets used to train and validate the potential.

The configuration space was chosen to cover the bonding environments that interstitial B and borides encounter in TiAl alloys. It comprises unary (Ti, Al, B), binary (Ti-Al, Ti-B, Al-B) and ternary Ti-Al-B structures taken from the Open Quantum Materials Database (OQMD) (19) and the Materials

Project (MP) (20), including both experimentally reported and theoretically predicted stable and metastable phases. Boron-specific configurations were added explicitly: TiAl compounds with interstitial B, and B-containing surfaces, vacancies, stacking faults and interfaces. The major Ti-B phases experimentally observed in TiAl alloys were also included, namely $TiB_2$ (C32), $Ti_3B_4$ (D7b), and the B27 and $B_f$ polymorphs of TiB (Table S1), together with interfaces between these compounds and the TiAl matrix, so that the potential is fitted across the boride-matrix boundary as well.

Finite-temperature bulk and defect configurations were sampled from *ab initio* molecular dynamics (AIMD) simulations between 300 and 2000 K, and the resulting snapshots were combined with the statically relaxed structures to form the initial dataset. This dataset was then expanded by active learning: configurations were selected on the basis of model uncertainty by the farthest point sampling method, labelled with the DFT settings above, and returned to training [Figure 1(a)]. After iteration, the final dataset combines physically constructed and adaptively sampled configurations, comprising a total of 7,316 structures. Detailed distributions of the structures across the subsystems are listed in Table S2.

### 2.2 NEP framework and training

Interatomic interactions in the Ti-Al-B system were described with the neuroevolution potential (NEP) framework (21) as implemented in the GPUMD package (22). In NEP, the total energy is written as a sum of atomic site energies, and each site energy is produced by a single-hidden-layer feedforward neural network acting on a rotationally invariant descriptor of the local atomic environment:

$$U_i = \sum_{\mu=1}^{N_{\text{neu}}} \omega_{\mu}^{(1)} \tanh\left( \sum_{\nu=1}^{N_{\text{des}}} \omega_{\mu\nu}^{(0)} q_{\nu}^{i} - b_{\mu}^{(0)} \right) - b^{(1)}. \quad (1)$$

Here, $q_{\nu}^{i}$ denotes the components of the rotationally invariant descriptor vector of atom $i$. $\omega^{(0)}$ and $\omega^{(1)}$ are the connection weights, while $b^{(0)}$ and $b^{(1)}$ represent the bias vectors. The descriptor vector contains both radial and angular components. Its radial part sums the contributions of all neighboring atoms $j$ within a cutoff distance $r_c^R$, where each pair contribution is expanded in Chebyshev polynomials through an element-dependent radial function $g_n(r_{ij})$, so that different atomic species pairs are distinguished explicitly:

$$q_n^i = \sum_{j \neq i} g_n(r_{ij}),\ 0 \leq n \leq n_{\max}^{\text{R}}. \quad (2)$$

The angular components extend the descriptor with three-body and four-body correlations expressed through spherical harmonics $Y_{lm}$. For the three-body contribution, the angular descriptor is written as:

$$q_{nl}^{i} = \sum_{m=-l}^{l} (-1)^{m} A_{nlm}^{i} A_{nl(-m)}^{i}, \tag{3}$$

$$A_{nlm}^{i} = \sum_{j \neq i} g_{n}(r_{ij}) Y_{lm}(\theta_{ij}, \phi_{ij}). \tag{4}$$

In these expressions, the spherical harmonics encode the angular arrangement of neighboring atoms relative to the central atom $i$. The four-body angular descriptor follows the formulation of Fan et al. (22), with $l_{\max}^{4b} = 2$. The index $n$ controls the radial resolution, while $l$ determines the angular momentum channel. The NEP parameters were optimized using the separable natural evolution strategy (SNES) (23) by minimizing a loss function that penalizes deviations from first-principles reference data.

For robustness and training efficiency, the empirical short-range repulsive Ziegler-Biersack-Littmark (ZBL) potential (24) was incorporated into NEP in a hybrid scheme that suppresses unphysical atomic clustering and reduces the risk of catastrophic failure in long MD runs. Training proceeded in two stages to improve the energy and virial accuracy while keeping the force fit stable. The first stage used energy, force and virial weights of 1.0, 1.0 and 0.5 with a batch size of 2000; the second raised the energy and virial weights to 5.0 and 2.0, kept the force weight unchanged, and increased the batch size to 8000. Detailed training parameters are listed in Table S3. NEP in GPUMD achieved 0.602 ns/day for a 1.05-million-atom model on a single RTX 4090 GPU, approximately 6.5 times faster than MEAM in LAMMPS on 40 CPU cores (Supplementary Note S1 and Figure S1).

### 2.3 Hybrid GCMC/MD simulations

Hybrid grand-canonical Monte Carlo/molecular dynamics (GCMC/MD) simulations were performed using the trained NEP model to determine the equilibrium incorporation behavior of B atoms in TiAl alloys. The simulations were carried out in the $\mu VT$ ensemble at $T = 900$ K, with the B chemical potential, system volume, and temperature fixed. The GCMC algorithm was implemented as an in-house module in the GPUMD package, enabling efficient sampling of B exchange with an external reservoir during MD simulations.  For the bulk, vacancy-containing, and interface models, 50 B insertion/deletion trials were attempted every 20 MD steps. For the polycrystalline models, 100 trials were attempted every 10 MD steps.

B insertion and deletion are attempted with equal probability, with deletion candidates selected uniformly from the existing B atoms. The insertion and deletion acceptance probabilities follow the standard grand-canonical Monte Carlo formulation (25):

$$P_{\text{ins}} = \min\left[1, \frac{V}{(N_{\text{B}}+1)\Lambda_{\text{B}}^3} e^{-\beta(\Delta\tilde{U}_{\text{ins}}-\mu_{\text{B}})}\right], \qquad P_{\text{del}} = \min\left[1, \frac{N_{\text{B}}\Lambda_{\text{B}}^3}{V} e^{-\beta(\Delta\tilde{U}_{\text{del}}+\mu_{\text{B}})}\right], \tag{5}$$

where $\beta = 1/(k_{\text{B}}T)$, $V$ is the cell volume, $N_{\text{B}}$ is the B atom count before the trial, and $\Lambda_{\text{B}} = h/\sqrt{2\pi m_{\text{B}} k_{\text{B}} T}$ is the thermal de Broglie wavelength of B. Energy changes are evaluated as final minus initial values using $\tilde{U} = U_{\text{NEP}} - N_{\text{B}} e_{\text{B}}^{\text{iso}}$, with $e_{\text{B}}^{\text{iso}} = -0.13$ eV, which corresponds to the isolated B atom energy. Insertion positions are sampled uniformly within the cell and redrawn if any periodic minimum-image interatomic distance is below 1.5 Å; the acceptance test is applied only after this criterion is satisfied, with the attempt abandoned after 1,001 unsuccessful placements. Although the 1.5 Å minimum-distance constraint formally violates detailed balance by introducing a bias in insertion attempts without a corresponding bias in deletion attempts, its practical impact on the simulation results is negligible.

### 2.4 Polycrystalline tensile simulations

To examine the effect of B on high-temperature deformation, a six-grain polycrystalline TiAl model with lamellar $\gamma/\gamma/\alpha_2$ units was constructed using Voronoi tessellation in Atomsk (26). Each $\gamma$ lamella was approximately 3.48 nm thick, and each $\alpha_2$ lamella was approximately 2.33 nm thick. The 26 nm cubic cell contained approximately 1.08 million atoms, with equivalent grain diameters of 17-19 nm chosen to minimize potential inverse Hall-Petch effects (27). Grain geometry and lamellar thicknesses were fixed across compositions to isolate B effects. After energy minimization and thermal equilibration, GCMC/MD sampling was performed at 900 K and a fixed B chemical potential of -7.5 eV for a cumulative MD duration of 4 ns, reaching a final overall B concentration of 1.163 at.%. Configurations with nominal B concentrations of 0.2, 0.5, 0.8, and 1.0 at.% were extracted at intermediate sampling stages, and a B-free reference was also included. All five models were equilibrated before tensile loading. In the initial tensile configurations of the four B-containing models, 76.72-95.85% of B atoms are isolated, while non-isolated B atoms predominantly form dimers (Figure S2). All five models underwent uniaxial tension at 900 K and an engineering strain rate of $2 \times 10^8$ $s^{-1}$ to a final engineering strain of ~20%.

## 3. Development and validation of the NEP-TiAlB

### 3.1 Training accuracy

Against the DFT reference data, the NEP model demonstrates high fidelity in reproducing energies, forces and virials with RMSEs of 4.56 meV/atom, 103.35 meV/Å and 25.26 meV/atom, respectively [Figures 1(b-d)]. The coverage of the dataset is assessed by projecting the high-dimensional descriptor vectors onto a two-dimensional space using principal component analysis (PCA) (28). The configurations are broadly distributed in this projection (Figure S3), which indicates that the dataset spans a wide range of chemical compositions and local atomic environments.

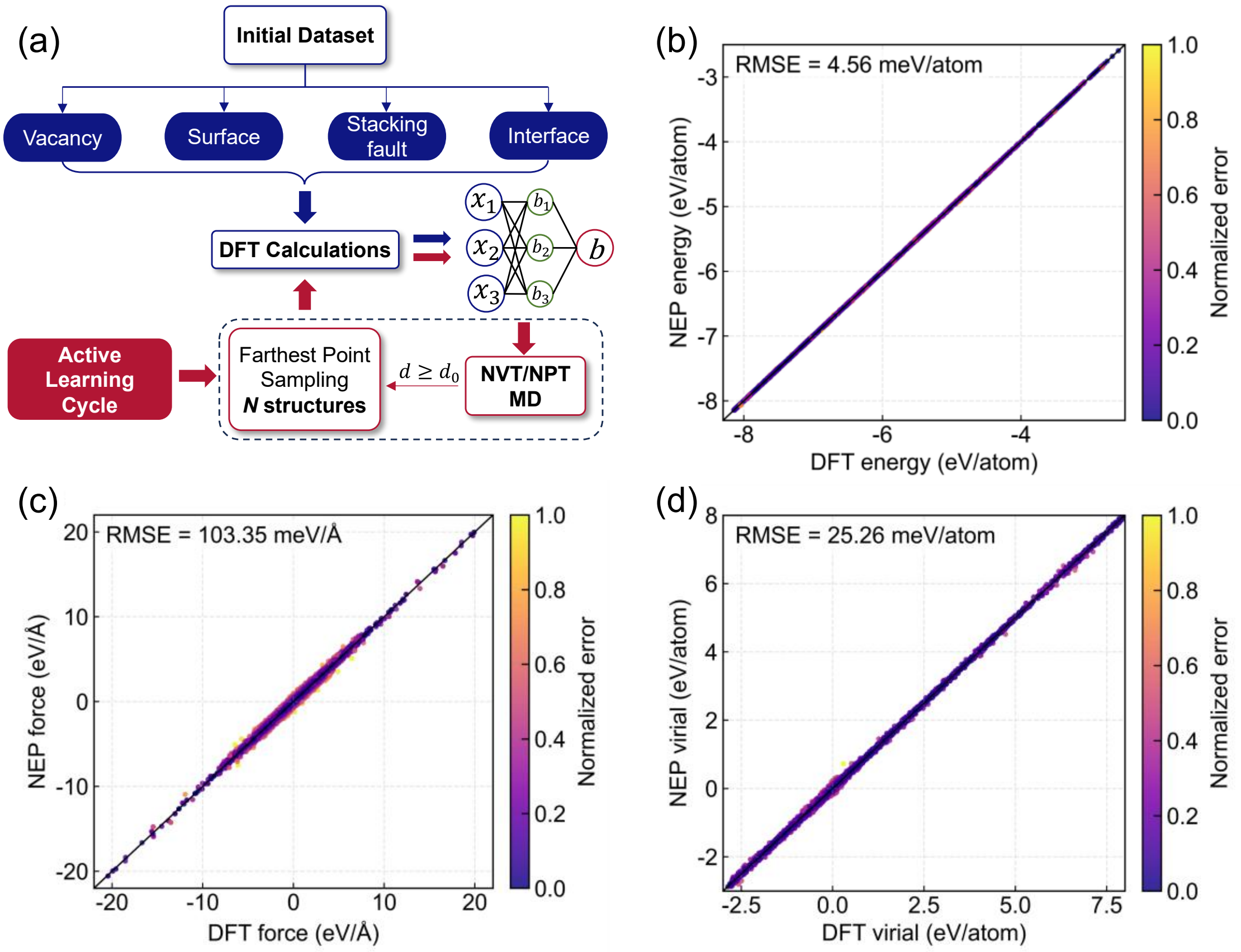

**Figure 1.** (a) Workflow for developing the NEP-TiAlB model using an active learning scheme. (b) Energies, (c) forces, and (d) virials as compared with the DFT reference values for the training dataset.

### 3.2 Basic bulk-phase properties

This NEP accuracy carries over to the bulk phases. The lattice parameters of $\gamma$-TiAl, $\alpha_2$-$Ti_3Al$, TiB (B27), TiB ($B_f$), $TiB_2$ (C32), and $Ti_3B_4$ ($D7_b$) agree with the available experimental data (29-34), with deviations from DFT values (35) within about 1% [Figure 2(a)]. Values for the six phases are listed in Table S4, and those for hcp-Ti and fcc-Al in Table S5. The corresponding MEAM predictions (36, 37) are included for comparison of lattice parameters and volumes per atom [Figure 2(b)]. The agreement extends beyond equilibrium geometries. Phonon dispersions predicted by NEP reproduce the experimental and DFT results, with no imaginary frequencies in any phase examined (Figure S4), which supports their vibrational properties and dynamical stability. The temperature-dependent lattice parameters of $\gamma$-TiAl and $\alpha_2$-$Ti_3Al$ also agree with the available experimental, DFT, EAM and MEAM data (Figure S5). The equation-of-state curves, which probe the energetics of volumetric deformation, follow the DFT results for four Ti-B structures and two B-doped TiAl configurations [Figures 2(c-h)]. The corresponding curves for the B-free $\gamma$-TiAl, $\alpha_2$-$Ti_3Al$ and D0$_{22}$-$TiAl_3$ are given in Figure S6.

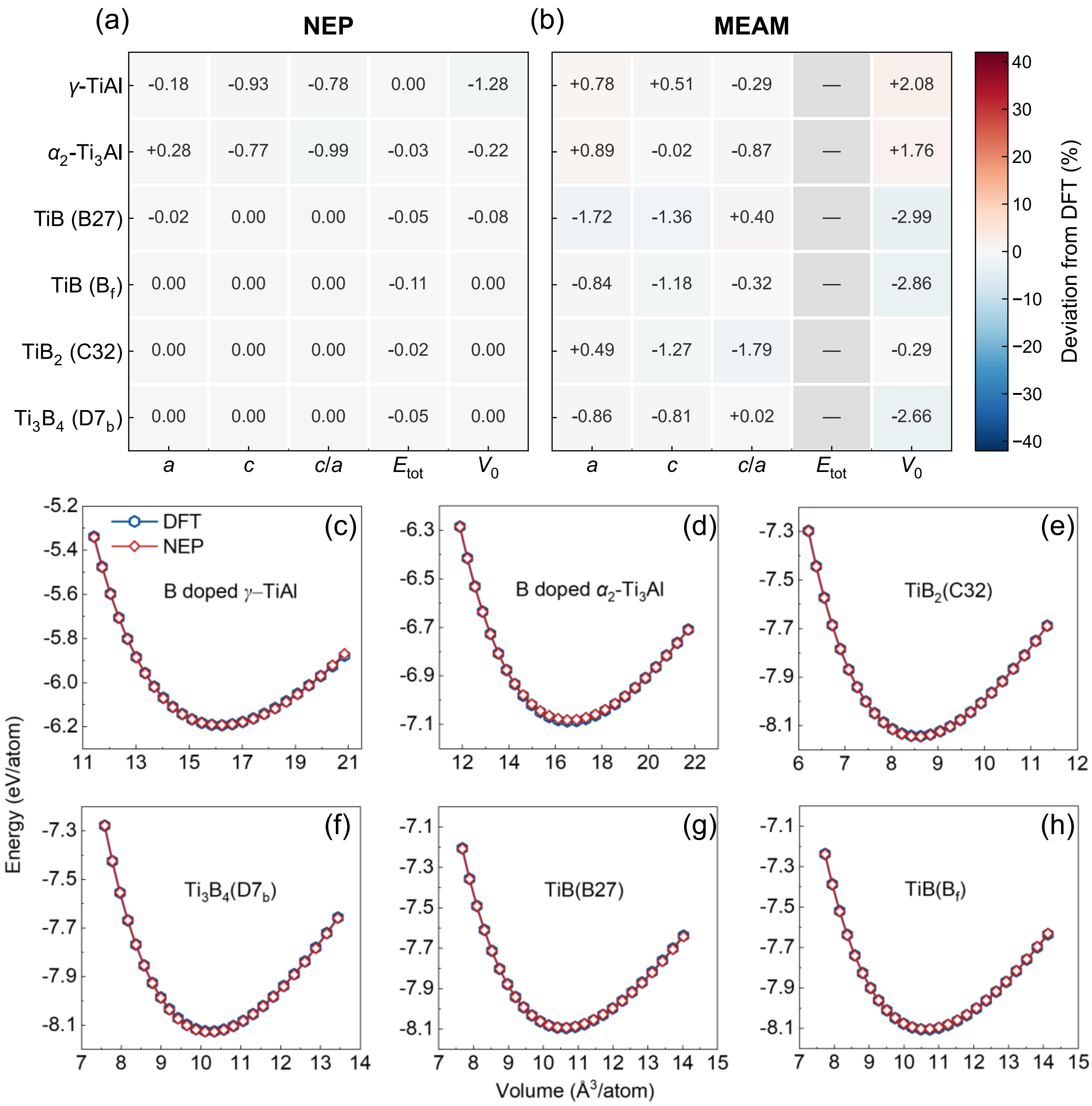

**Figure 2.** Validation of bulk-phase properties predicted by the NEP-TiAlB model. (a,b) Percentage deviations in lattice constants ($a$, $c$, and $c/a$) and volumes per atom ($V_0$) predicted by NEP (a) and MEAM (b) relative to the corresponding DFT values. Total energy per atom ($E_{tot}$) is compared with DFT only for NEP. Dashes indicate that total energy comparisons are omitted for MEAM because the energy references differ. (c-h) Equation-of-state (EOS) curves for (c) B-doped $\gamma$-TiAl, (d) B-doped $\alpha_2$-$Ti_3Al$, (e) $TiB_2$ (C32), (f) $Ti_3B_4$ (D7$_b$), (g) TiB (B27), and (h) TiB (B$_f$), obtained from NEP and DFT calculations. The B-doped $\gamma$-TiAl and $\alpha_2$-$Ti_3Al$ models use 49-atom supercells ($Ti_{24}Al_{24}B$ and $Ti_{36}Al_{12}B$, respectively), each containing one interstitial B atom at a concentration of 2.04 at.%. The lattice vectors are uniformly scaled within a linear strain range of ± 10% to generate the EOS data.

The same level of agreement holds for the elastic response (Table S6). The elastic constants of the Ti-Al and Ti-B compounds agree well with DFT and experiment, whereas MEAM shows larger deviations and even nonphysical values such as a negative $C_{66}$ for TiB (B27). Values for hcp-Ti, fcc-Al, $D0_{22}$-TiAl3, TiB ($B_f$) and $Ti_3B_4$ are listed in Table S7. The bulk modulus $B$, shear modulus $G$, Young's modulus $E$ and Poisson ratio $v$ were derived with the Voigt-Reuss-Hill method (38), and the resulting effective polycrystalline properties also agree well with the reference values (Table S8).

The deviations remain small across compounds and elastic quantities, as the parity plot in Figure 3 shows: the NEP results lie close to the diagonal, whereas the MEAM results scatter more widely. Expressed as mean absolute percentage errors (MAPE), the NEP values are 3.2-7.1% for the four phases benchmarked against DFT, compared with 15.6-57.0% for MEAM. The temperature-dependent elastic constants of $\gamma$-TiAl and $\alpha_2$-$Ti_3Al$ are validated in Figures S7 and S8, confirming the accuracy of the NEP model for finite-temperature mechanical predictions.

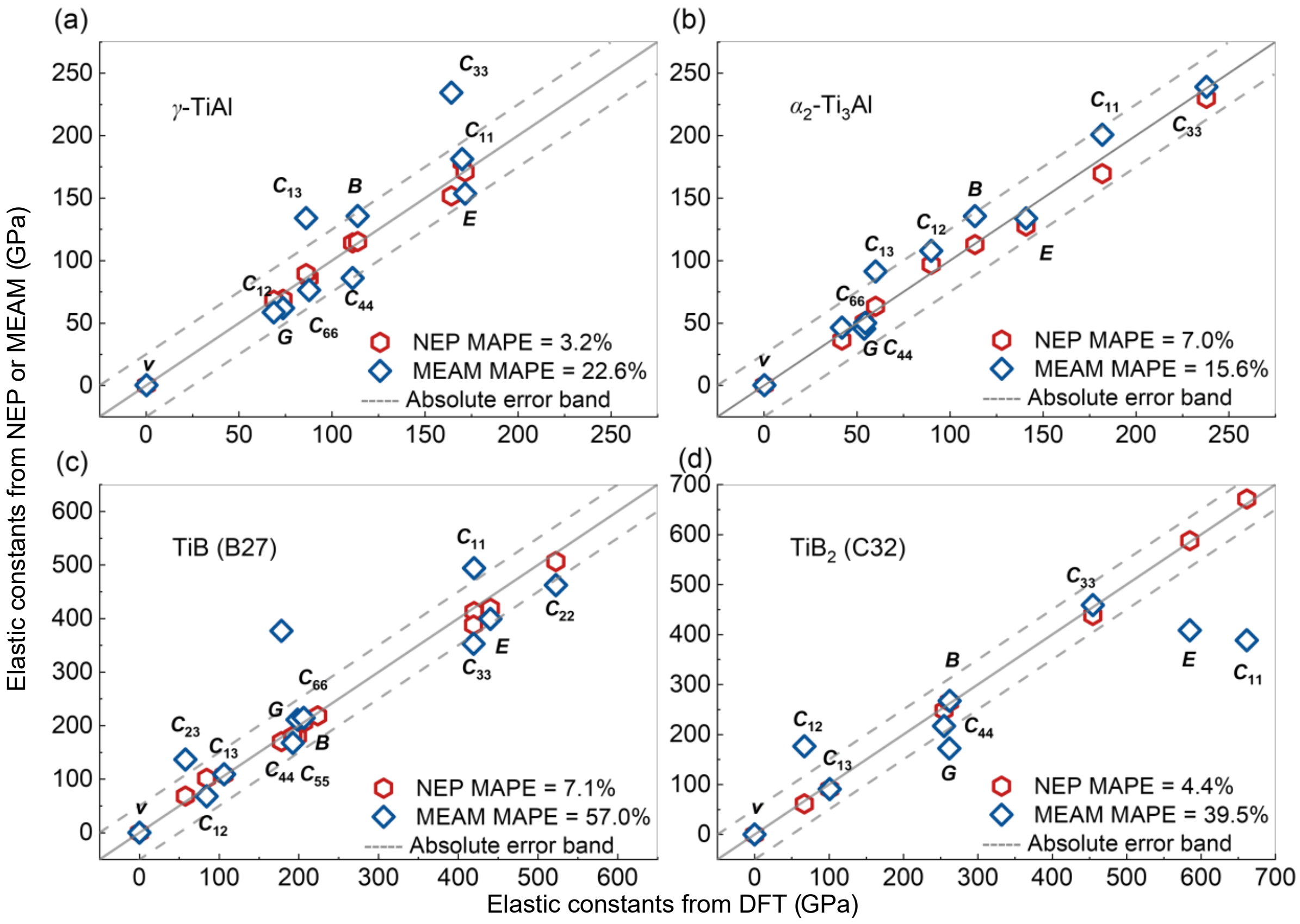

**Figure 3.** Comparison of the elastic constants predicted by NEP and MEAM with DFT reference values for (a) $\gamma$-TiAl, (b) $\alpha_2$-$Ti_3Al$, (c) TiB (B27), (d) $TiB_2$ (C32). MAPEs are calculated relative to DFT results.

The elastic constants further distinguish the two families of phases. The Ti-B compounds have larger principal elastic constants ($C_{11}$, $C_{22}$, $C_{33}$) and higher $B$, $G$ and $E$ than the Ti-Al phases, indicating a stronger resistance to elastic deformation and a higher intrinsic stiffness. Their Poisson ratios are at most 0.161, well below the $v$ = 1/3 threshold that separates ductile from brittle behavior (39). The brittleness is therefore more pronounced in the Ti-B compounds. The Ti-Al phases lie below or close to this threshold. Ti-B phases can therefore act as stiff but brittle reinforcing constituents in TiAl alloys, whose volume fraction, morphology and spatial distribution must be controlled (40).

### 3.3 B-related defect and interface energetics

#### 3.3.1 Polar and non-polar surface energies

Surface energy controls faceting, roughening, crystal growth and surface segregation in alloys (41), and contributes to the interfacial cleavage energy that governs damage in TiAl. Surfaces in the Ti-Al-B system fall into two classes: non-polar surfaces, whose energies are independent of the chemical potential, and polar Ti-B surfaces, whose energies depend on surface termination and chemical environments. All NEP surface energies listed in Table S9 deviate from the corresponding DFT values by less than 4%. However, MEAM errors are much larger, typically 20-60%, which limits its transferability for these surfaces.

For Ti-B compounds, however, some low-index surfaces are polar and their surface compositions may deviate from the bulk stoichiometry depending on the surface termination. In such cases, surface energy $\sigma$ depends on the chemical potentials of Ti and B, and is defined as follows:

$$\sigma = \frac{1}{2A}\left(E_{\text{slab}} - N_{\text{Ti}}\mu_{\text{Ti}}^{\text{slab}} - N_{\text{B}}\mu_{\text{B}}^{\text{slab}} + PV - TS\right), \tag{6}$$

where $A$ is the surface area, $E_{\text{slab}}$ is the slab energy, and $\mu_{\text{Ti}}$ and $\mu_{\text{B}}$ are the chemical potentials of Ti and B, respectively; the $PV$ and $TS$ terms are neglected. $N_{\text{Ti}}$ and $N_{\text{B}}$ are the numbers of Ti and B atoms in the slab, respectively. Taking $TiB_2$ as an example, the surface is assumed to be in equilibrium with the bulk compound, making the two chemical potentials interdependent. We therefore express them as deviations from their respective bulk reference values:

$$\Delta\mu_{\text{Ti}} = \mu_{\text{Ti}}^{\text{slab}} - \mu_{\text{Ti}}^{\text{bulk}}, \Delta\mu_{\text{B}} = \mu_{\text{B}}^{\text{slab}} - \mu_{\text{B}}^{\text{bulk}} \tag{7}$$

Combining the equilibrium condition with the formation enthalpy of $TiB_2$ gives:

$$\Delta\mu_{\text{Ti}} + 2\Delta\mu_{\text{B}} = \Delta H_f(\text{TiB}_2). \tag{8}$$

Stability requires $\Delta\mu_{\text{Ti}} \leq 0$ and $\Delta\mu_{\text{B}} \leq 0$, which restricts $\Delta\mu_{\text{B}}$ to

$$\frac{1}{2}\Delta H_f(\text{TiB}_2) \leq \Delta\mu_{\text{B}} \leq 0, \Delta H_f(\text{TiB}_2) < 0. \tag{9}$$

TiB follows the same treatment with its 1:1 stoichiometry, giving $\Delta H_f$ (TiB) $\leq \Delta\mu_B \leq 0$ (see Supplementary Note S2 for the detailed derivation).

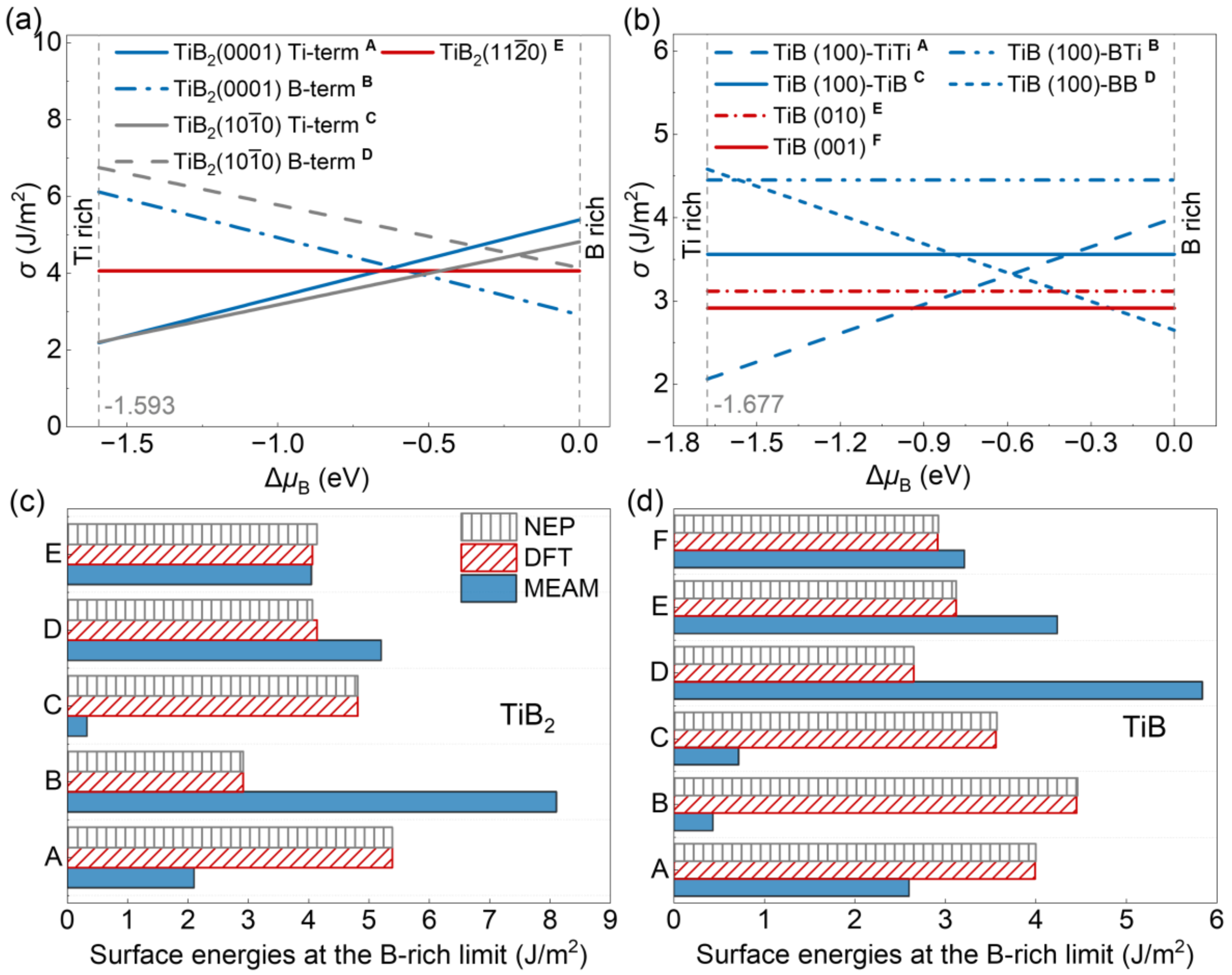

**Figure 4.** Relaxed surface energies of $TiB_2$ (C32) and TiB (B27) surfaces. (a, b) DFT-calculated surface energies of low-index $TiB_2$ (a) and TiB (b) surfaces as functions of the B chemical-potential difference $\Delta\mu_B$. (c-d) Comparison of surface energies predicted by NEP and MEAM with DFT reference values at the B-rich limit for $TiB_2$ (c) and TiB (d), respectively.

Figures 4(a) and 4(b) show the DFT surface energies of $TiB_2$ (C32) and TiB (B27) as functions of $\Delta\mu_B$ based on the above derivations. The allowed ranges of $\Delta\mu_B$ are -1.593 to 0 eV and -1.677 to 0 eV, respectively. These are formal bounds that neglect competing borides, rather than thermodynamic stability limits. For $TiB_2$, Ti-terminated (0001) and (10-10) surfaces are more stable under Ti-rich conditions, whereas B-terminated surfaces are favored at the B-rich limit, which is consistent with previous DFT reports (42). The (11-20) surface is stoichiometric and non-polar, with a constant surface energy. For TiB, the polar (100) surface has four terminations, labelled by their outermost layer: TiTi-, BTi-, TiB- and BB-terminated. The TiB- and BTi-terminated surfaces are stoichiometric, with energies independent of $\Delta\mu_B$, whereas the TiTi- and BB-terminated surfaces are non-stoichiometric. The (001) and (010) surfaces are stoichiometric as well. Among the four TiB (100) terminations, the TiTi-terminated surface is favored under Ti-rich conditions ($\Delta\mu_B$ < -0.58 eV); the BB-terminated surface becomes more stable towards the B-rich limit. At the B-rich limit, the stability decreases in the order TiB (100)-BB, TiB (001), TiB (010), TiB (100)-TiB, TiB (100)-TiTi and TiB (100)-BTi. This ordering agrees with earlier DFT calculations on TiB (43). Termination stability in both borides is therefore set by the chemical environment rather than by the plane itself.

Figures 4(c) and 4(d) further compare the NEP and MEAM surface energies with DFT at the B-rich limit. NEP reproduces both the absolute surface energies and the relative stability trends. However, MEAM (37) deviates for several surfaces and overestimates the B-terminated $TiB_2$ (0001) surface by about 5 $J/m^2$. This error reverses the stability of that surface relative to the other terminations. Together with the non-polar surfaces, these results establish that NEP reproduces the surface energetics required by the interfacial and cleavage calculations below.

### 3.3.2 Stacking fault energies

Boron alters the generalized stacking fault energy (GSFE) of TiAl alloys and thereby influences dislocation nucleation and plastic deformation. To test whether the potential captures this effect, Figures 5(a-c) compare the GSFE curves of B-doped $\gamma$-TiAl and $\alpha_2$-$Ti_3Al$ calculated using NEP and DFT, with one interstitial B atom placed near the slip plane in the doped structures. For $\gamma$-TiAl, the GSFE curves are calculated along a four-segment displacement path on the {111} plane, with B at Ti-rich or Al-rich octahedral interstitial sites (OIS). The $\gamma$-line paths pass through the critical superlattice intrinsic stacking fault (SISF), the antiphase boundary (APB) and the complex stacking fault (CSF) [Figures 5(a) and 5(b)]. For $\alpha_2$-$Ti_3Al$, the $\gamma$ line is computed on the basal {0001} plane along the $[01\bar{1}0]$ direction, with B at a Ti-rich OIS [Figure 5(c)]. The arrows indicate the displacement increments along each path. For the GSFE calculations, the cell vectors were fixed, and atomic relaxation was allowed only normal to the slip plane.

Figures 5(d) and 5(e) summarize the characteristic stacking fault energies predicted by NEP for the B-free and B-doped systems, while the corresponding DFT results are shown in Figures S9(c) and S9(d). The DFT results show that the B-induced changes depend on the phase and the fault configuration. For $\gamma$-TiAl with B at the Ti-rich OIS, $\gamma_{USISF}$ decreases, whereas $\gamma_{UCSF}$ and $\gamma_{CSF}$ increase and $\gamma_{SISF}$ changes only slightly. For $\alpha_2$-$Ti_3Al$, B does not uniformly lower the characteristic energies: $\gamma_{UAPB}$ and $\gamma_{SISF}$ increase. NEP predicts opposite trends for several quantities, including $\gamma_{USISF}$ and $\gamma_{CSF}$ in Ti-rich $\gamma$-TiAl and $\gamma_{UAPB}$ in $\alpha_2$-$Ti_3Al$, and overestimates the increase in $\gamma_{SISF}$ in Ti-rich $\gamma$-TiAl.

Overall, NEP reasonably reproduces the characteristic stacking fault energies and the general shapes of the DFT-predicted GSFE curves for B-doped $\gamma$-TiAl and $\alpha_2$-$Ti_3Al$. Most deviations in the unstable stacking fault energies are within 15%, although some discrepancies remain in chemically sensitive configurations. In particular, the predicted trends for certain small B-induced energy changes differ from DFT, including $\gamma_{USISF}$ and $\gamma_{CSF}$ in Ti-rich $\gamma$-TiAl and $\gamma_{UAPB}$ in $\alpha_2$-$Ti_3Al$. These results demonstrate that NEP provides a reasonable description of the overall stacking fault energetics, despite some differences in capturing subtle energy variations induced by dilute B. The corresponding NEP and DFT results for pristine TiAl systems are provided in Figure S9 for comparison.

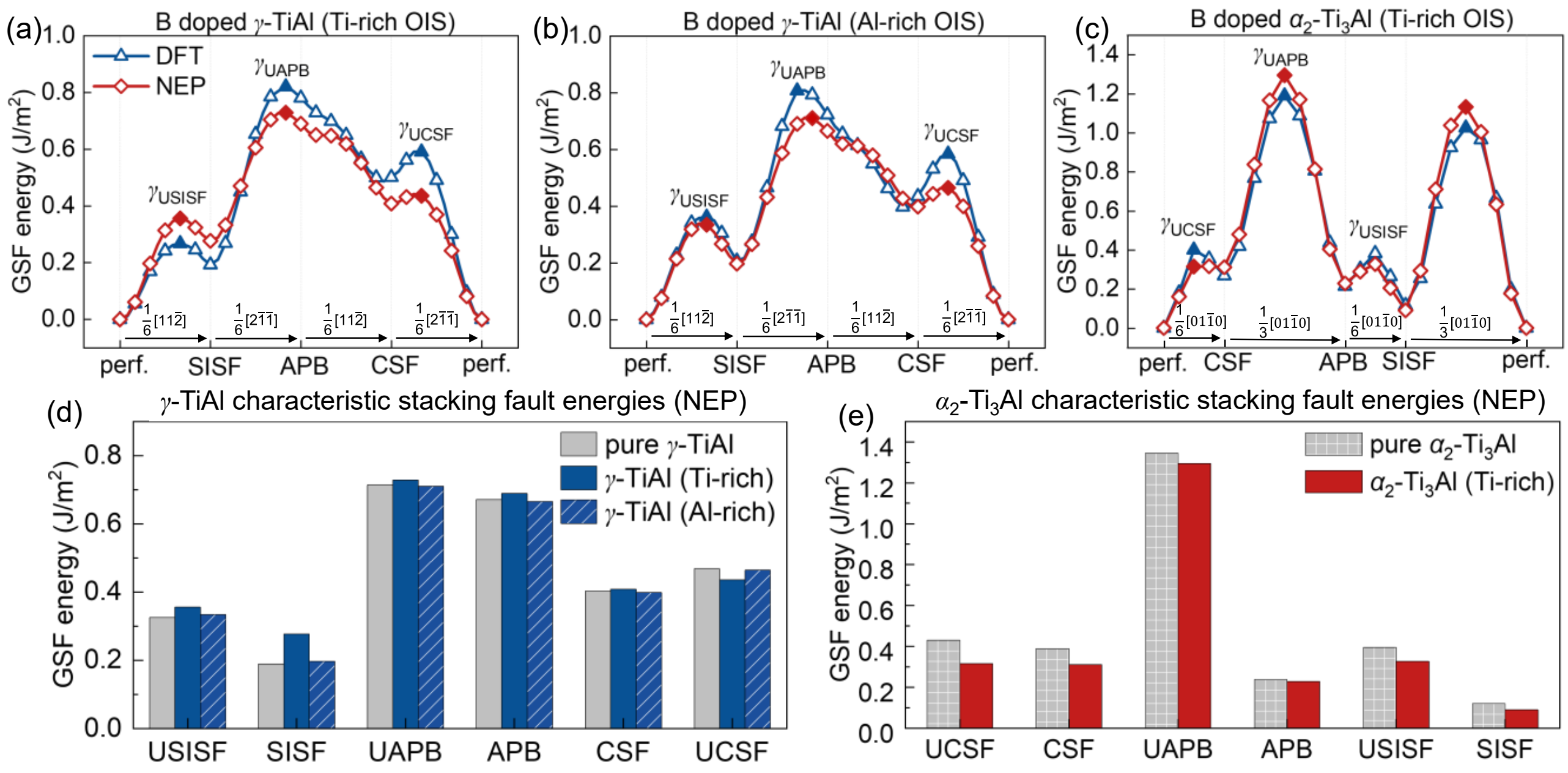

**Figure 5.** Generalized stacking fault energy (GSFE) curves of $\gamma$-TiAl and $\alpha_2$-$Ti_3Al$ with B concentration of 2 at.% predicted by NEP and DFT. (a-b) GSFE profiles along the complete four-partial dissociation path of a $[10\bar{1}]$ super dislocation in $\gamma$-TiAl with B at Ti-rich and Al-rich OISs, respectively. (c) GSFE profile along the straight $[01\bar{1}0]$ path O → CSF → APB → SISF → O′ in $\alpha_2$-$Ti_3Al$ with B at a Ti-rich OIS. $\gamma_{UCSF}$, $\gamma_{UAPB}$, and $\gamma_{USISF}$ denote the GSFE maxima along the O-CSF, CSF-APB, and APB-SISF segments. Arrows indicate the displacement increments. (d-e) Comparison of characteristic stacking fault energies in $\gamma$-TiAl and $\alpha_2$-$Ti_3Al$ calculated by NEP. Gray, blue, and red spheres denote Ti, Al, and B atoms, respectively, consistent throughout this work.

### 3.3.3 Adhesion of TiAl/$TiB_2$ interfaces

$TiB_2$ ceramic particles are widely used as effective reinforcements in TiAl-based composites because of their high strength, high melting point, as well as good chemical and thermal compatibility with $\gamma$-TiAl (44, 45). Their addition improves the mechanical properties of the matrix, and this gain depends on the bonding across the TiAl/$TiB_2$ interface (46). To further assess the transferability of the NEP model to intermetallic/ceramic heterogeneous interfaces, the adhesion properties of the TiAl/$TiB_2$ interface are examined and compared with DFT results.

The interface models were constructed using the experimentally (47-49) established $TiB_2$ (0001)/$\gamma$-TiAl (111) orientation [Figures 6(a) and 6(b)]. Two $TiB_2$ (0001) terminations, Ti- and B-terminated, are combined with six stacking sequences of the $\gamma$-TiAl (111) surface, giving twelve models. The bonding strength is quantified by the work of adhesion $W_{ad}$, the reversible work per unit area needed to separate the interface into two free surfaces:

$$W_{ad} = \frac{(E_{TiB_2} + E_{TiAl} - E_{TiB_2/TiAl})}{A}, \tag{10}$$

where $E_{TiB_2}$ and $E_{TiAl}$ are the total energies of the separated slabs, respectively. $E_{TiB_2}/E_{TiAl}$ is the total energy of the relaxed interface system, and $A$ is the interfacial area.

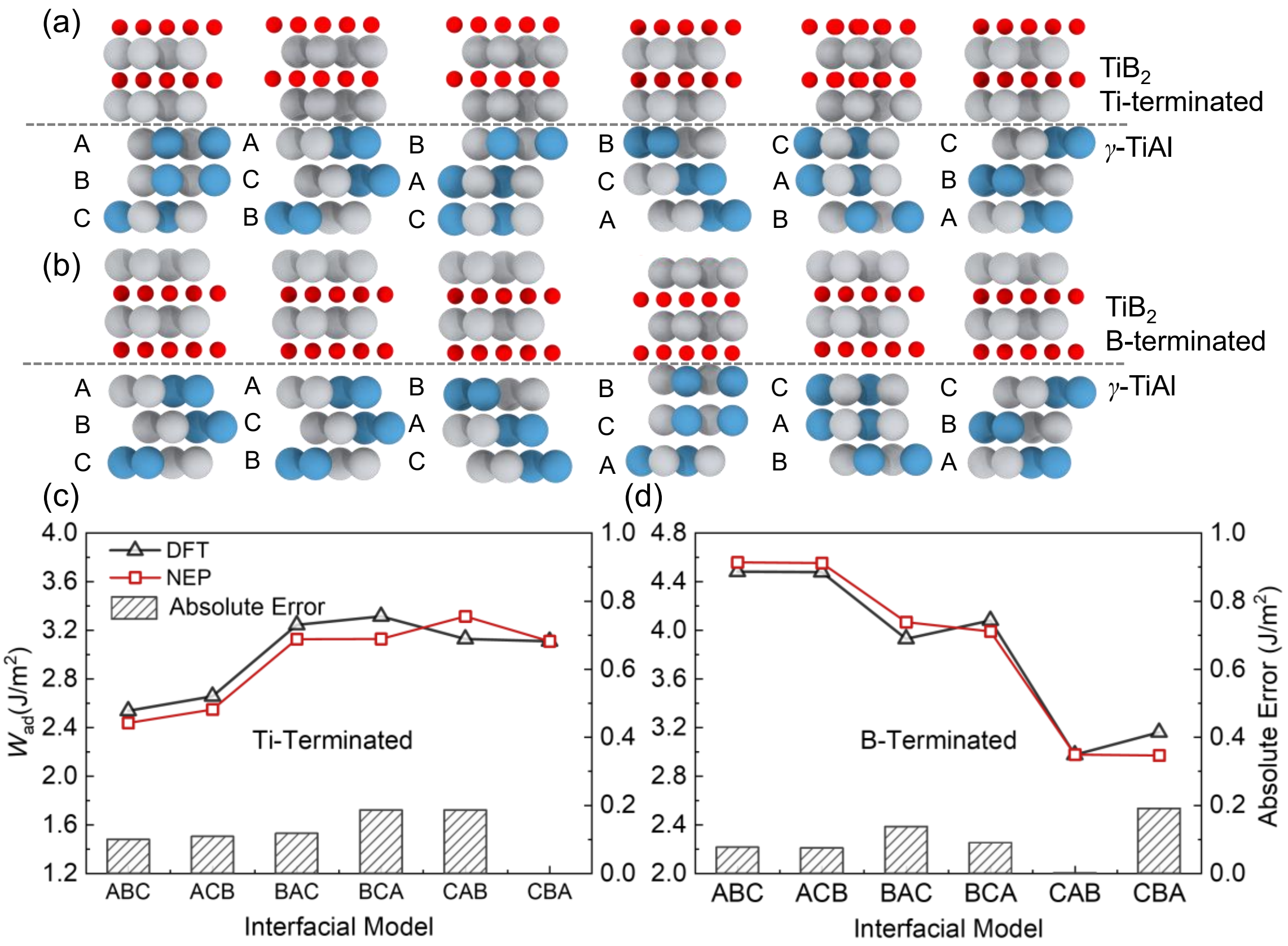

**Figure 6.** Work of adhesion for twelve $TiB_2$ (0001)/TiAl (111) interface models. (a-b) Ti-terminated and B-terminated interfaces with six stacking sequences: ABC, ACB, BAC, BCA, CAB, and CBA. (c-d) Corresponding $W_{ad}$ values for the Ti-terminated and B-terminated interfaces, respectively.

Figures 6(c) and 6(d) compare the work of adhesion from NEP and DFT for the Ti- and B-terminated interfaces. In both panels, a larger $W_{ad}$ indicates stronger interfacial bonding. NEP reproduces the DFT values to within about 0.2 J/m$^2$ and follows the stacking-dependent adhesion trends. For Ti-terminated interfaces, the $W_{ad}$ values are lower, and the ABC and ACB stackings adhere most weakly. By contrast, these stackings are the most strongly bound when the interface is B-terminated, whereas CAB and CBA are the weakest. The preferred stacking therefore reverses with the termination of the $TiB_2$ surface. This reversal, correctly reproduced by NEP, is essential for describing how borides bond to the TiAl matrix.

## 4. Defect-mediated redistribution of B in TiAl alloys

### 4.1 Thermodynamic and kinetic trapping of B

Before considering defect- and interface-mediated redistribution, we establish the thermodynamic and kinetic behavior of B in bulk $\gamma$-TiAl and $\alpha_2$-$Ti_3Al$. We first identify its preferred occupation sites and migration pathways, and then quantify how the B chemical potential controls equilibrium incorporation in each phase.

#### 4.1.1 Site preference

The site that B occupies governs its migration, vacancy trapping and interfacial uptake, so we first establish the preferred sites by formation energies. The formation energies $E_f$ of interstitial and substitutional B are defined as a single expression:

$$E_{\mathrm{f}} = E_{\text{Ti-Al+B}} - E_{\text{Ti-Al}} - \sum_i n_i \mu_i, \tag{11}$$

where $E_{\text{Ti-Al+B}}$ and $E_{\text{Ti-Al}}$ are the total energies of the TiAl phase with and without B, respectively. For interstitial B, $n_B = 1$; whereas for substitutional B, $n_B = 1$ and $n_X = -1$, where X denotes the replaced Ti or Al atom. The reference chemical potentials are taken from stable $\alpha$-boron ($B_{36}$) for B (50), hcp-Ti for Ti, and fcc-Al for Al. Using these reference states allows the formation energies of all candidate sites to be compared on a consistent basis.

**Table I.** DFT calculated formation energies $E_f$ (eV) of B occupying different interstitial sites (octahedral and tetrahedral) and substitutional sites in $\gamma$-TiAl and $\alpha_2$-$Ti_3Al$.

| | Octahedral interstitial sites | | | Tetrahedral interstitial sites | Substitutional sites | |
|---|---|---|---|---|---|---|
| | 6Ti | 2Al+4Ti | 4Al+2Ti | | $B_{Ti}$ | $B_{Al}$ |
| $\alpha_2$-$Ti_3Al$+B | -0.333 | 0.608 | — | 2.183 | 1.801 | 1.643 |
| $\gamma$-TiAl+B | — | 0.059 | 1.201 | 3.331 | 1.873 | 1.614 |

B strongly favors octahedral interstitial sites (OIS) over tetrahedral interstitial sites (TIS) and substitutional sites (SS) in both TiAl phases (Table I). The lowest $E_f$ occurs in Ti-rich octahedral environments, at -0.333 eV for the OIS (6Ti) in $\alpha_2$-$Ti_3Al$ and 0.059 eV for the OIS (4Ti+2Al) in $\gamma$-TiAl. In $\alpha_2$-$Ti_3Al$, B placed at a TIS relaxes spontaneously to a neighboring OIS. The preference follows from the larger electronegativity difference for B-Ti than for B-Al, which strengthens B-Ti bonds and stabilizes Ti-rich octahedral environments (16). B therefore preferentially occupies Ti-rich octahedral sites in both phases, providing the basis for the subsequent analyses of diffusion and incorporation.

### 4.1.2 Diffusion barriers and equilibrium B incorporation

B diffusion in TiAl alloys proceeds by thermally activated hopping between neighboring stable interstitial sites through saddle points on the potential energy surface. This barrier is the activation energy that governs the diffusion kinetics. We evaluate these barriers in $\gamma$-TiAl and $\alpha_2$-$Ti_3Al$ with the climbing-image nudged elastic band (CI-NEB) method (51). Intermediate images are generated between the initial and final configurations of each path, and the minimum energy path is optimized until the force falls below 0.01 eV/Å.

Figures 7(a) and 7(b) show the resulting diffusion barriers, which depend strongly on the local chemical environment and on the migration path. B preferentially occupies Ti-rich octahedral sites, the 4Ti+2Al site in $\gamma$-TiAl and the 6Ti site in $\alpha_2$-$Ti_3Al$. In $\gamma$-TiAl, migration between two 4Ti+2Al sites has the highest barrier of 3.2 eV, whereas paths between Al-rich 2Ti+4Al sites require only about 1.1 eV. The step from a 4Ti+2Al to a 2Ti+4Al site still costs about 2.3 eV. These pathways suggest that B diffusion is controlled by escape from the preferred Ti-rich sites, with an effective activation energy of approximately 2.3 eV, close to the experimental value of 2.07 eV reported by Divinski et al. (52). In $\alpha_2$-$Ti_3Al$, migration from the 6Ti site to a 4Ti+2Al site costs about 2.5 eV, following the same pattern. Ti-rich octahedral environments therefore act as effective traps for B, combining the lowest formation energies with the highest escape barriers. NEP reproduces the DFT barriers for every path considered, which is the property required by the hopping-time estimates at 900 K in Section 4.4.

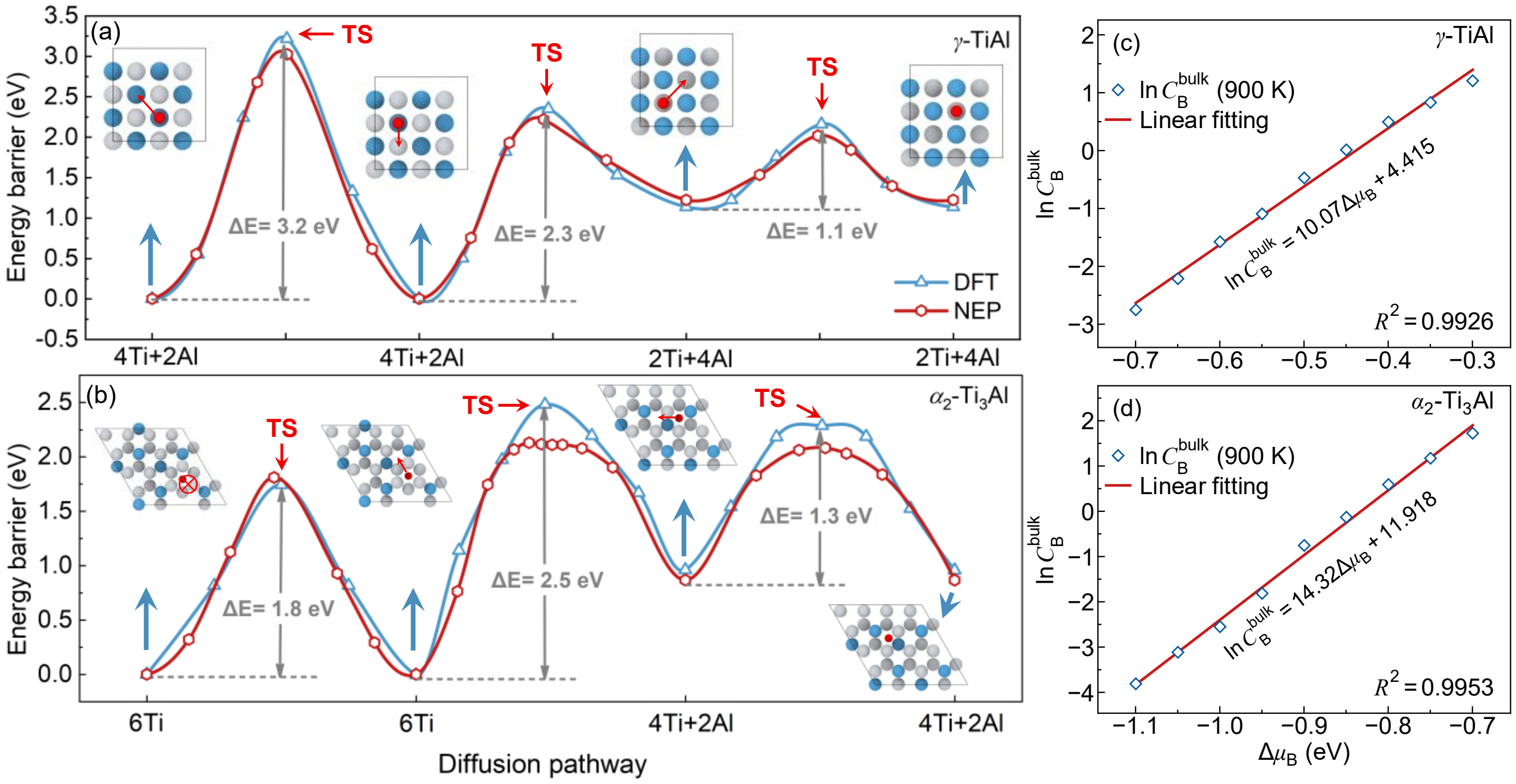

**Figure 7.** (a-b) Diffusion energy barriers for B migrating between different interstitial sites. $\gamma$-TiAl: 4Ti+2Al → 4Ti+2Al → 2Ti+4Al → 2Ti+4Al. $\alpha_2$-$Ti_3Al$: 6Ti → 6Ti → 4Ti+2Al → 4Ti+2Al. (c-d) ln C versus $\Delta\mu_B$ for $\gamma$-TiAl and $\alpha_2$- $Ti_3Al$ at 900 K from GCMC/MD simulations, with $C$ in at.% and solid lines indicating linear fits. Here, $\Delta\mu_B = \mu_B - \mu_B^0$, where $\mu_B^0$ = -6.7003 eV/atom is the NEP $\alpha$-B reference energy per atom.

To examine how the B chemical potential affects B incorporation in TiAl alloys, a series of GCMC/MD simulations are performed at 900 K over a range of B chemical potentials, $\mu_B$. These simulations relate $\mu_B$ to the bulk B concentration $C_B^{bulk}$, thereby providing the thermodynamic basis for selecting appropriate $\mu_B$ used in the vacancy-mediated incorporation and interfacial segregation simulations below.

Figures 7(c) and 7(d) show that ln$C$ varies approximately linearly with $\Delta\mu_B$ in both phases, indicating an approximately exponential dependence of B concentration on chemical potential over the sampled ranges. The fitted slopes are 10.07 and 14.32 $eV^{-1}$ for $\gamma$-TiAl and $\alpha_2$-$Ti_3Al$, respectively, with corresponding $R^2$ values of 0.9926 and 0.9953. At a B concentration of 1 at.%, the fitted chemical potential for $\gamma$-TiAl is approximately 0.39 eV higher than that for $\alpha_2$-$Ti_3Al$. This offset is close to the DFT formation-energy difference of 0.392 eV between the preferred interstitial sites in the two phases (Table I), consistent with the more favorable incorporation of B into $\alpha_2$-$Ti_3Al$.

### 4.2 Vacancy-assisted B trapping

#### 4.2.1 B-vacancy interactions

Interstitial B interacts with vacancies in TiAl alloys, modifying defect stability and local relaxation even at dilute concentrations. Thus, the complementary quantities $E_{bind}$ (B-vacancy binding energy) and $E_{relax}$ (relaxation energy), calculated using DFT and NEP, describe this interaction. $E_{bind}$ measures the tendency of B and a vacancy to form a complex rather than remain isolated, with negative values indicating energetically favorable association. For a vacancy on the $X$ site, with $X$ = Ti or Al,

$$E_{\text{bind}}(B - V_X) = [E(V_X, \text{B}) + E(\text{host})] - [E(\text{host} + B) + E(V_X)], \quad (12)$$

where $E$(host) and $E$(host+B) are the energies of the defect-free and B-doped TiAl supercells, respectively. $E(V_X)$ and $E(V_X$, B) are the energies of the supercells containing the $X$ vacancy alone and the $X$ vacancy with B. $E_{relax}$ quantifies the energy reduction caused by atomic relaxation and is defined as the energy difference between the unrelaxed and relaxed B-containing vacancy configurations:

$$E_{\text{relax}} = E_{\text{unrelaxed}}(V_X, B) - E_{\text{relaxed}}(V_X, B). \quad (13)$$

Here, $E_{unrelaxed}$ and $E_{relaxed}$ are the total energies before and after atomic relaxation, respectively. A larger $E_{relax}$ indicates a greater energy reduction through atomic rearrangement in the B-containing vacancy configuration, including relaxation of the surrounding Ti-Al atoms and possible migration of B into the vacancy site.

Table II summarizes the predicted binding energies of the B-$V_{Ti}$ and B-$V_{Al}$ complexes in both phases from NEP and DFT calculations. Both methods place the stable complexes preferentially at $V_{Al}$, notably in $\alpha_2$-$Ti_3Al$. NEP reproduces the DFT results and ordering for most cases. Figure 8 shows that the relaxation behavior is also vacancy- and phase-dependent. For $V_{Ti}$, B is not trapped in $\gamma$-TiAl [Figure 8(a)] but relaxes into the vacancy site in $\alpha_2$-$Ti_3Al$ [Figure 8(b)]. For $V_{Al}$, B occupies the vacancy site in both phases and forms stable complexes [Figures 8(c) and 8(d)]. The relaxation energies in Figure 8(e) further support these trends, where "near" and "far" denote the initial position of B relative to the vacancy. In $\gamma$-TiAl, $V_{Ti}$-near shows a smaller energy reduction upon relaxation than $V_{Ti}$-far, whereas $V_{Al}$-near shows a substantially greater reduction. In $\alpha_2$-$Ti_3Al$, both near-vacancy configurations exhibit greater energy reductions than their far counterparts, particularly for $V_{Al}$. Overall, NEP reproduces the vacancy- and phase-dependent relaxation trends predicted by DFT, despite some quantitative discrepancies for the $V_{Al}$-near configuration in $\alpha_2$-$Ti_3Al$. Both methods consistently indicate that B preferentially associates with Al vacancies.

**Table II.** Solute-vacancy binding energy, $E_{bind}$ (eV), of B-$V_{Ti}$ and B-$V_{Al}$ complexes in $\gamma$-TiAl and $\alpha_2$-$Ti_3Al$, calculated using NEP and DFT.

| Structure | Sites | DFT | NEP |
|---|---|---|---|
| B doped $\gamma$-TiAl | B-$V_{Ti}$ | 0.261 | 0.216 |
| | B-$V_{Al}$ | -1.275 | -1.158 |
| B doped $\alpha_2$-$Ti_3Al$ | B-$V_{Ti}$ | -1.092 | -1.073 |
| | B-$V_{Al}$ | -2.694 | -1.684 |

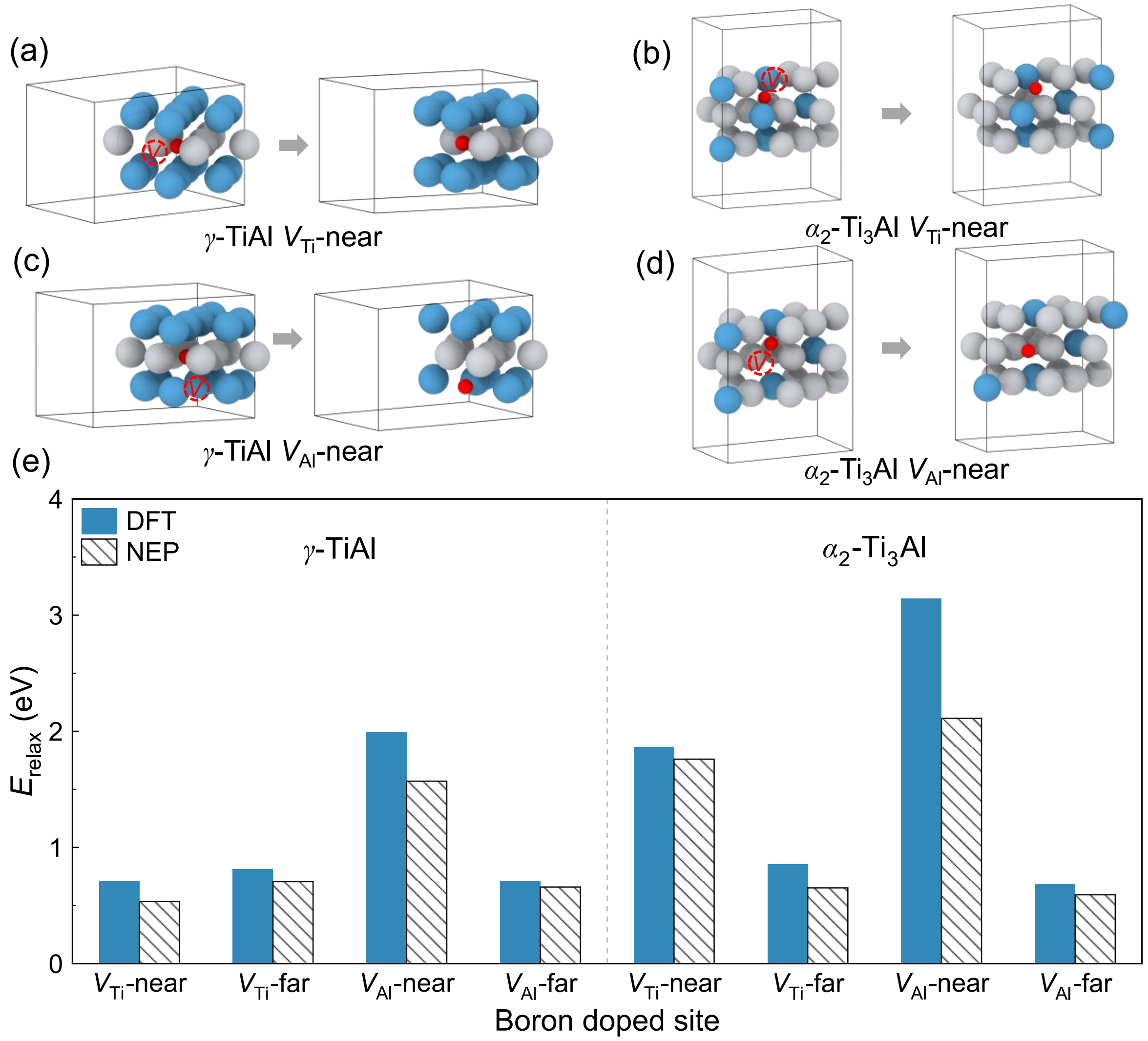

**Figure 8.** Structural relaxation and relaxation energies ($E_{relax}$) of Ti and Al vacancies in B-doped TiAl phases. Vacancies are initially placed near or far from interstitial B, with $V_{Ti}$ and $V_{Al}$ denoting Ti and Al vacancies, respectively. Initial and relaxed configurations are shown for $V_{Ti}$ (a,b) and $V_{Al}$ (c,d) in $\gamma$-TiAl (a,c) and $\alpha_2$-$Ti_3Al$ (b,d), with B initially near each vacancy. (e) Comparison of $E_{relax}$ from NEP and DFT for both vacancy types near and far from B.

### 4.2.2 Vacancy-mediated B incorporation

TiAl alloys are susceptible to creep during high-temperature service, with vacancy-mediated processes such as dislocation climb and atomic diffusion contributing to the deformation rate. Vacancies therefore play a key role in high-temperature deformation and solute redistribution in these alloys. To investigate vacancy-mediated B incorporation, GCMC/MD simulations are performed for $\gamma$-TiAl and $\alpha_2$-$Ti_3Al$ containing Ti vacancies, Al vacancies, or no vacancies. The effects of B chemical potential, applied stress, and vacancy concentration on equilibrium B uptake and local enrichment are subsequently examined. Details of the vacancy-containing models are provided in Supplementary Note S3.

Figures 9(a-d) show the time-averaged vacancy-B radial distribution functions (RDFs) in $\gamma$-TiAl and $\alpha_2$-$Ti_3Al$, characterizing the local enrichment of B around Ti and Al vacancies. The vacancy-B RDF is calculated as

$$g_{V-\mathrm{B}}(r_k) = \frac{\left\langle N_\mathrm{B}(r_k^- \le d_{V-\mathrm{B}} < r_k^+)\right\rangle}{\Delta V_k \left\langle N_\mathrm{B}^\mathrm{total} / V\right\rangle}. \tag{14}$$

Here, $N_\mathrm{B}$ denotes the number of B atoms within the radial shell bounded by $r_k^-$ and $r_k^+$, $d_{V\text{-B}}$ is the vacancy-B distance, and $\Delta V_k$ is the shell volume. $N_\mathrm{B}^\mathrm{total}$ and $V$ denote the total number of B atoms and the simulation cell volume, respectively. The angle brackets indicate time averaging over the sampled configurations. Values of ( $g_{V\text{-B}}(r_k) > 1$ ) indicate local B enrichment relative to the average B number density. The RDF profiles show more pronounced near-vacancy peaks around Al vacancies than Ti vacancies, indicating stronger local B accumulation and a greater trapping tendency at Al vacancies. This preference is consistent with the B-vacancy binding energies discussed in Section 4.2.1, further supporting the important role of Al vacancies in B incorporation.

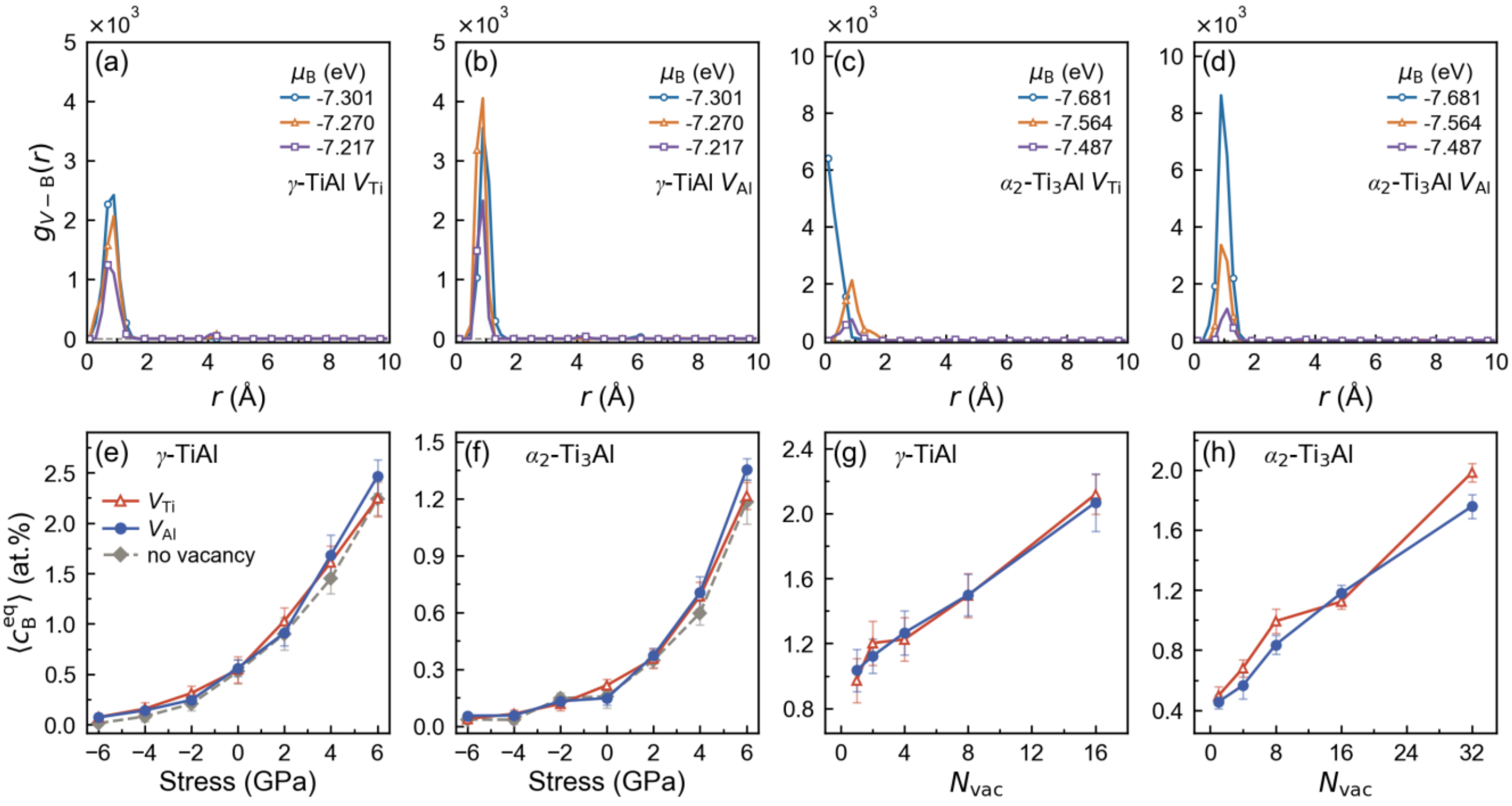

**Figure 9.** Vacancy-mediated B incorporation and enrichment in $\gamma$-TiAl and $\alpha_2$-$Ti_3Al$ obtained from GCMC/MD simulations at 900 K. (a-d) Normalized radial distributions $g_{V-\mathrm{B}}(r)$ around Ti and Al vacancies in $\gamma$-TiAl and $\alpha_2$-$Ti_3Al$. The B number density is calculated in spherical shells of width 0.2 Å and normalized by the mean B number density $\langle N_\mathrm{B}^\mathrm{total} / V\rangle$ of each vacancy-containing system. The curves represent averages over 100 frames sampled during the final 5,000 MD steps. (e-f) Average

equilibrium B concentration, $\langle c_{\mathrm{B}}^{\mathrm{eq}} \rangle$, as a function of applied stress. (g-h) Average equilibrium B concentration, $\langle c_{\mathrm{B}}^{\mathrm{eq}} \rangle$, as a function of vacancy number, $N_{\mathrm{vac}}$.

Given the mechanical loads experienced by TiAl alloys during service, we further examined the effect of applied stress on the average equilibrium B concentration, $<c_{\mathrm{B}}^{\mathrm{eq}}>$, around a monovacancy [Figures 9(e) and 9(f)]. $<c_{\mathrm{B}}^{\mathrm{eq}}>$ increases progressively as the applied stress changes from compression to tension in both phases. Tensile stress increases the local free volume, reducing the energetic cost of accommodating B at interstitial sites or near vacancies. This stress dependence is more pronounced in $\gamma$-TiAl than in $\alpha_2$-$Ti_3Al$. Over the examined stress range, differences between vacancy types are smaller than the stress-induced changes, suggesting that applied stress has a greater influence on B incorporation than vacancy type. Increasing vacancy concentration further promotes B incorporation into the lattice. Figures 9(g) and 9(h) show that $<c_{\mathrm{B}}^{\mathrm{eq}}>$ increases with vacancy number in both phases, consistent with additional free volume and energetically favorable trapping sites provided by vacancies. B incorporation and local enrichment in TiAl alloys are therefore jointly controlled by B chemical potential, applied stress, and vacancy concentration. Vacancies serve as preferential sites for B accumulation during high-temperature service, thereby influencing solute redistribution and defect evolution.

### 4.3 Interface-Boron interactions

#### 4.3.1 B-mediated interfacial cohesion

The high density of lamellar $\gamma/\gamma$ and $\gamma/\alpha_2$ interfaces in TiAl alloys strongly influences their mechanical response. Their distinct chemical environments and excess free volume may also favor interstitial B trapping. To assess B site preference and interfacial cohesion, the formation energy ($E_{\mathrm{f}}$) and cleavage energy ($\gamma_{\mathrm{cl}}$) are calculated for different interstitial sites. The formation energy of interstitial B is defined as:

$$E_{\mathrm{f}} = E_{I\text{-B}} - E_I - E_{\mathrm{B}},\ I = \gamma/\gamma \text{ or } \gamma/\alpha_2, \tag{15}$$

where $I$ denotes the $\gamma/\gamma$ or $\gamma/\alpha_2$ interface, $E_{I\text{-B}}$ and $E_I$ are the supercell energies with and without B, and $E_{\mathrm{B}}$ is the energy per atom of $\alpha$-B. The interfacial cleavage energy $\gamma_{\mathrm{cl}}$ represents the energy per unit area required to separate a solid into two semi-infinite halves, creating two free surfaces without diffusion or plastic deformation (16):

$$\gamma_{\mathrm{cl}\,(\gamma/\alpha_2\text{-B})} = \frac{E_{\mathrm{slab1}} + E_{\mathrm{slab2}} - E_{\mathrm{interface}}}{A}, \tag{16}$$

where $A$ is the interfacial area, $E_{\mathrm{interface}}$ is the energy of the B-containing interface, and $E_{\mathrm{slab1}}$ and $E_{\mathrm{slab2}}$ denote the energies of the two separated slabs. B remains in its original slab position during cleavage.

Figures 10(a) and 10(b) illustrate six interstitial sites at each interface, covering Ti-rich and Al-rich environments in the upper bulk-like, interfacial, and lower bulk-like regions. Details of the interface models are provided in Supplementary Note S3. Figures 10(c-f) present the corresponding $E_f$ and $\gamma_{cl}$ values. At the $\gamma/\gamma$ interface [Figures 10(c) and 10(e)], Ti-rich sites (S1, S3, S5) have lower formation energies than Al-rich sites (S2, S4, S6). More importantly, interfacial S3 and S4 sites yield higher cleavage energies than bulk-like sites. The NEP cleavage energies for B occupying S3 and S4 exceed that of the B-free TT interface, indicating that B occupation at these sites increases the work required to separate the interface. Among all sites, site S3 combines a relatively low $E_f$ with the highest $\gamma_{cl}$, making its Ti-rich environment most favorable for B trapping and interfacial strengthening. At the $\gamma/\alpha_2$ interface [Figures 10(d) and 10(f)], $\gamma_{cl}$ exhibits weaker site dependence, whereas formation energies still favor Ti-rich sites (S1′, S3′, S5′). The NEP cleavage energies at all sampled sites are slightly higher than that of the B-free interface. Overall, NEP reproduces the DFT trends in both $E_f$ and $\gamma_{cl}$, capturing the preference of B for Ti-rich environments and the site-dependent changes in cleavage energy.

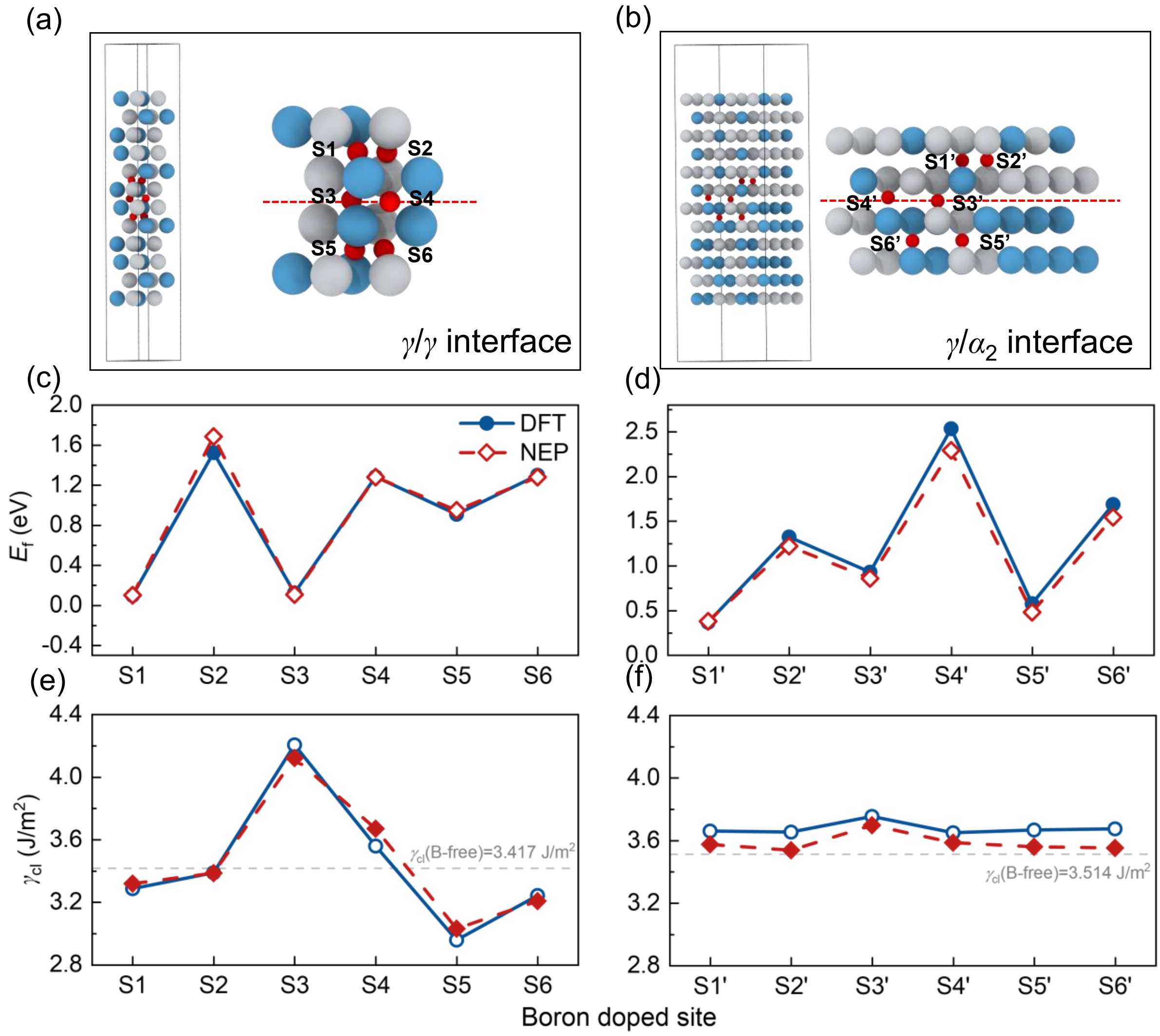

**Figure 10.** B-doped interface structures and corresponding formation energies $E_f$ and cleavage energies $\gamma_{cl}$ calculated by DFT and NEP. (a) $\gamma/\gamma$ true-twin interface with six representative interstitial B sites: S1/S5 (Ti-rich bulk-like), S2/S6 (Al-rich bulk-like), S3 (Ti-rich interfacial), and S4 (Al-rich interfacial). (b) $\gamma/\alpha_2$ interface with six interstitial B sites: S1′ (6Ti) and S2′ (4Ti+2Al) in bulk-like $\alpha_2$; S3′ (Ti-rich) and S4′ (Al-rich) at the interface; and S5′ (Ti-rich) and S6′ (Al-rich) in bulk-like $\gamma$. Site-dependent B formation energies (c,d) and cleavage energies (e,f) for the $\gamma/\gamma$ true-twin (c,e) and $\gamma/\alpha_2$ (d,f) interfaces, respectively. Horizontal dashed lines indicate the NEP cleavage energies of the corresponding B-free interfaces.

#### 4.3.2 Interfacial segregation of B

To examine the equilibrium distribution of B at TiAl interfaces, we perform GCMC/MD simulations using representative $\gamma/\gamma$ and $\gamma/\alpha_2$ models. The $\mu_B$ is set to -7.4 eV for the true twin (TT), pseudo twin (PT), rotation boundary (RB), and coherent $\gamma/\alpha_2$ models. For the semi-coherent $\gamma/\alpha_2$ model, $\mu_B$ is reduced to -7.6 eV because the higher B concentration at -7.4 eV hindered stable equilibration in preliminary simulations. Accordingly, the following discussion focuses on the spatial distribution and segregation tendency of B rather than absolute B concentrations. Simulation and analysis details are provided in Supplementary Note S3.

Figures 11(a-c) show the B probability-density profiles along the $\gamma/\gamma$ interface normal for the TT, PT, and RB models, with both ends representing interfacial regions under periodic boundary conditions. B is nearly uniformly distributed across all $\gamma/\gamma$ models, without pronounced interfacial peaks. In contrast, the $\gamma/\alpha_2$ models exhibit B enrichment predominantly on the $\alpha_2$ side, consistent with the $\mu_B$-$C_B^{bulk}$ relationship [Figures 7(c) and 7(d)]. The coherent model shows preferential B partitioning into the $\alpha_2$ phase rather than localized interfacial segregation [Figure 11(d)], whereas the semi-coherent model exhibits a distinct interfacial B-enrichment peak [Figure 11(e)].

The stronger B segregation at the semi-coherent $\gamma/\alpha_2$ interface likely arises from defect-assisted trapping, where local strain, excess free volume, and potentially favorable Ti-rich sites promote B incorporation. This highlights the role of interface coherency and local defect structure in B segregation. Similar solute segregation at boundaries has also been reported for interstitial H in Fe-based alloys (53) and B in Ni-Cr alloys (54), with implications for interfacial cohesion and defect evolution. These findings distinguish preferential B partitioning into the $\alpha_2$ phase from localized enrichment at the semi-coherent $\gamma/\alpha_2$ interface.

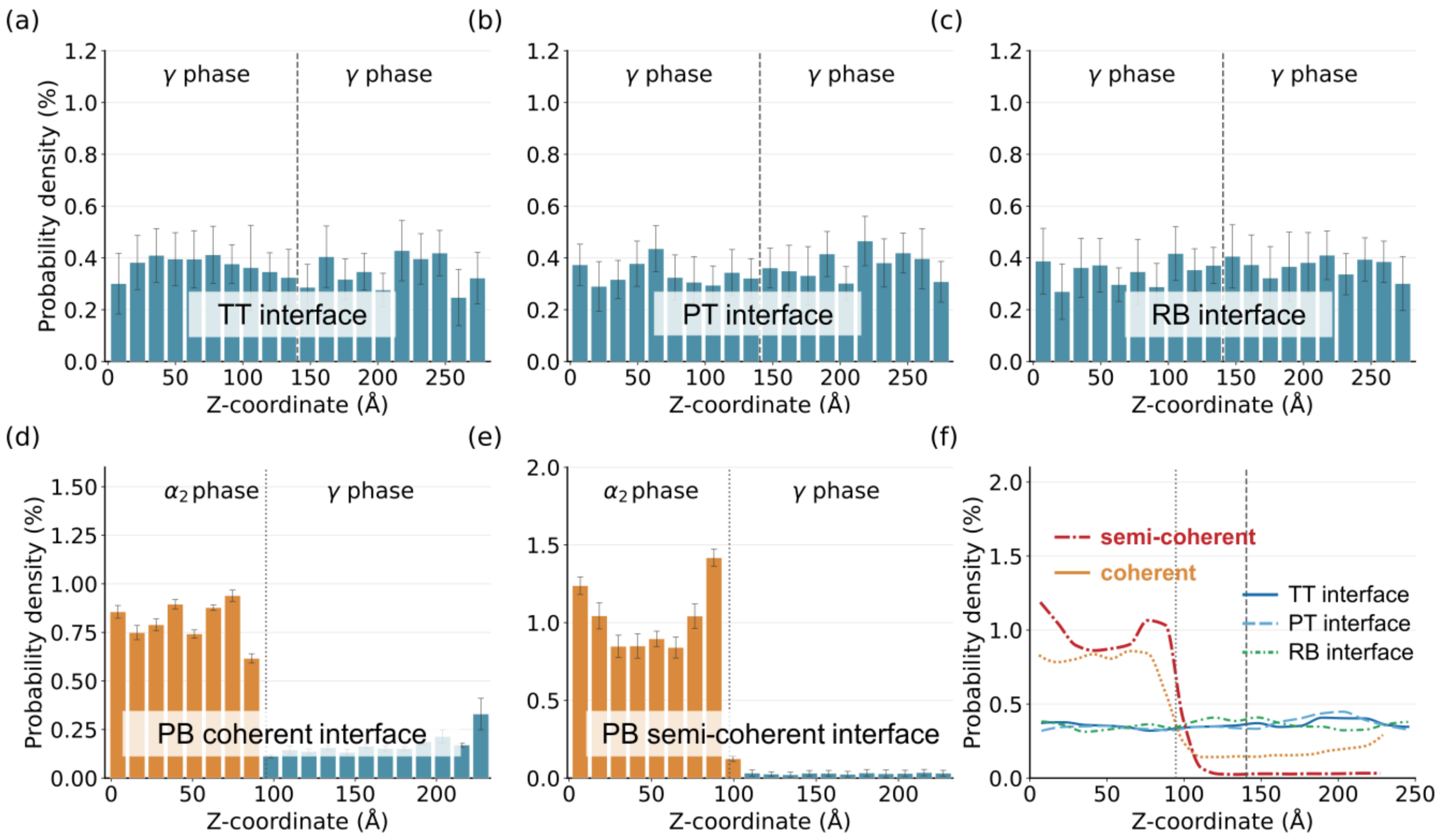

**Figure 11.** Probability density distributions of B along the interface-normal direction (*z* direction) in $\gamma/\gamma$ and $\gamma/\alpha_2$ models obtained from GCMC/MD simulations. (a-c) B distributions at the $\gamma/\gamma$ interfaces in the TT, PT, and RB models, respectively. (d-e) B distributions in the coherent and semi-coherent $\gamma/\alpha_2$ phase-boundary models. Blue and orange bars indicate B probability densities in the $\gamma$ and $\alpha_2$ regions, respectively. (f) Smoothed fitting probability density curves from (a-e), compared across $\gamma/\gamma$ and $\gamma/\alpha_2$ models.

### 4.4 B strengthening in TiAl polycrystals

To investigate B-induced strengthening, uniaxial tensile simulations are performed on six-grain polycrystalline TiAl models containing lamellar $\gamma/\gamma/\alpha_2$ structures with B concentrations ranging from 0 to 1.0 at.% at 900 K, as described in Section 2.4. For each B content, three tensile simulations are performed using the same initial atomic configuration but different initial velocities. Peak and average flow stresses are evaluated separately for each run and are reported as the mean ± sample standard deviation across the three runs. Figures 12(a-c) show that B addition enhances the deformation resistance of polycrystalline TiAl. Adding 1.0 at.% B increases the peak stress from 3.93 to 4.27 GPa and the average flow stress from 3.30 to 3.61 GPa. A similar strengthening trend has also been experimentally observed in Ti-43Al-6Nb-1Mo-1Cr-xB alloys (55), where increasing the B content up to 1 at.% enhanced the high-temperature tensile strength by approximately 121 MPa at 800°C. To quantify B enrichment within the structural regions identified at each strain, the regional B partition factor, $P_r$, is defined as:

$$P_r = \frac{N_{\mathrm{B}}^r / N_{\mathrm{B}}}{N_{\text{Ti-Al}}^r / N_{\text{Ti-Al}}}, \tag{17}$$

Here, $N_{\mathrm{B}}^r$ and $N_{\text{Ti-Al}}^r$ are the numbers of B and Ti-Al atoms in region $r$, while $N_{\mathrm{B}}$ and $N_{\text{Ti-Al}}$ are the corresponding system totals. Regions with $P_r > 1$ are B-enriched, those with $P_r = 1$ have the system-average B concentration, and those with $P_r < 1$ are B-depleted. Details of the calculation of $P_r$ are provided in the Supplementary Note S4.

Figures 12(d-f) show that $P_r$ for regions classified as $\gamma/\alpha_2$ phase boundaries increases from approximately 1.1 to 1.5 during tension, whereas $\gamma/\gamma$ interfaces approach the system-average distribution from an initially depleted state. In disordered regions, $P_r$ decreases from ~10.4 to ~6.6 but remains substantially above unity. At 1 at.% B, the fraction of B atoms in these regions increases from 91.2% to 93.0%, while the corresponding Ti-Al host fraction increases from 8.8% to 14.0%. Thus, the decrease in $P_r$ reflects region expansion rather than net B depletion. To assess whether B can escape from Ti-rich traps during tensile loading, mean escape times were estimated using transition state theory,

$$\tau = \nu_0^{-1} \exp\left(\frac{E_{\mathrm{m}}}{k_{\mathrm{B}}T}\right), \tag{18}$$

Here, we use an attempt frequency of $10^{13}$ $s^{-1}$ and the migration barriers in Figures 7(a) and 7(b). At 900 K, the trap-escape barriers of 2.3 and 2.5 eV in $\gamma$-TiAl and $\alpha_2$-$Ti_3Al$ correspond to mean escape times of approximately 0.8 and 10 s, respectively, far exceeding the approximately 1 ns tensile duration. Extensive bulk B diffusion is unlikely over the tensile timescale. Changes in $P_r$ are therefore attributed primarily to the evolution of the interfacial and disordered regions and their constituent atomic populations during deformation.

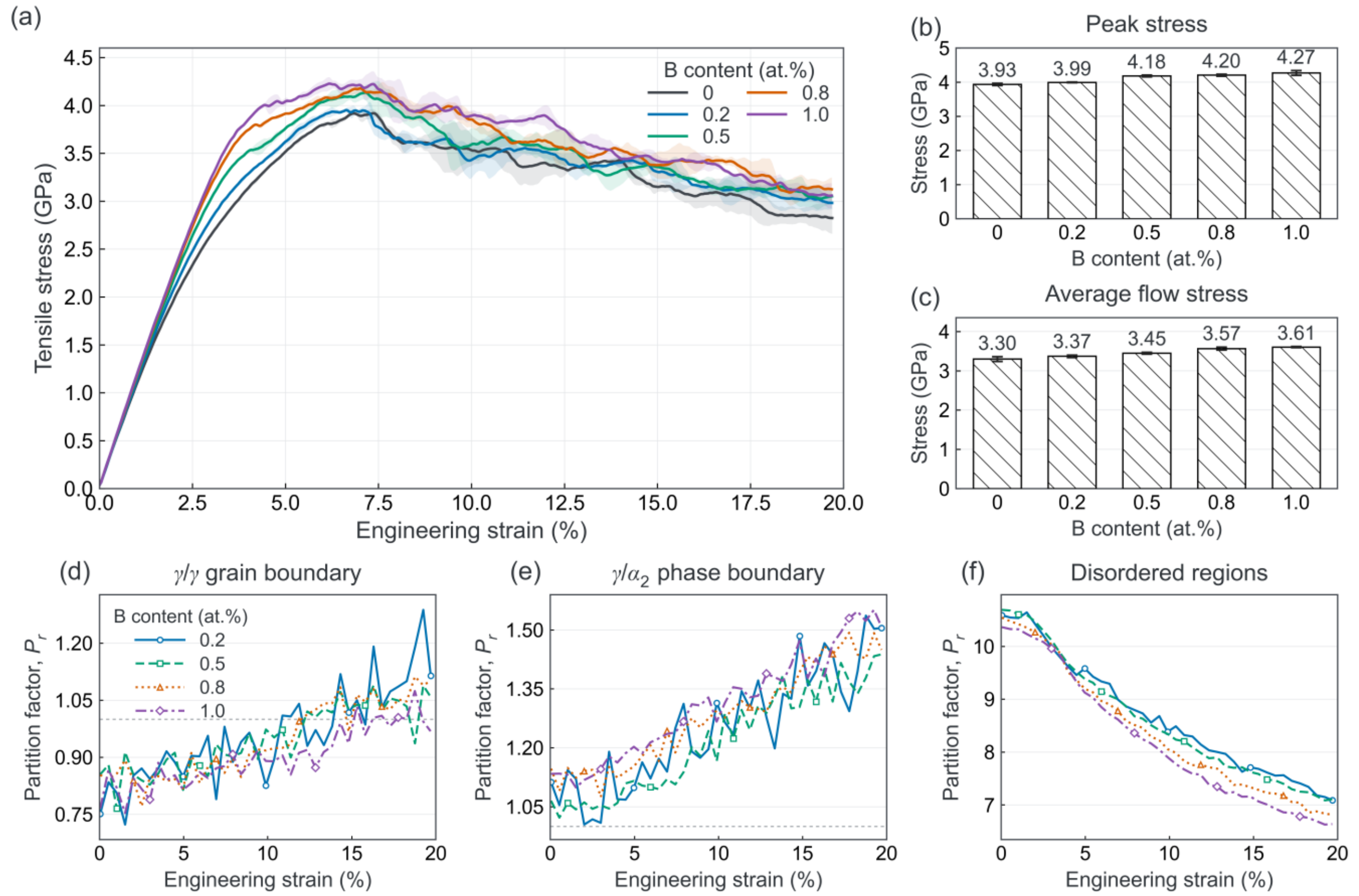

**Figure 12.** High-temperature tensile response and regional B enrichment in polycrystalline TiAl. (a-c) Stress-strain curves at 900 K, peak stress, and average flow stress over the post-peak strain interval from ~7% to 19.7%, respectively. (d-f) Evolution of the regional B partition factor for $\gamma/\gamma$ interfacial, $\gamma/\alpha_2$ phase-boundary, and disordered regions, respectively. The horizontal dashed lines in (d,e) indicate $P_r = 1$.

The enhanced tensile resistance of B-doped TiAl is accompanied by B enrichment in disordered regions and near grain boundaries, together with reduced local shear-strain accumulation. Figure 13(a) shows the microstructural evolution at five tensile strains, revealing B accumulation primarily near grain boundaries and the progressive development of fault/twin-related lamellae within grains. Figure 13(b) shows that regions classified as Other by PTM exhibit higher mean local shear strain than FCC/HCP regions. Compared with B-free TiAl at the same tensile strain, 1 at.% B reduces the mean local shear strain in Other regions by approximately 42%, 23%, and 9% at strains of 4%, 7.2%, and 19.7%, respectively. These observations support a possible strengthening mechanism in which B restricts local atomic rearrangements in disordered regions near grain boundaries.

B also influences intragranular fault and twin evolution. At 19.7% strain, fewer fault/twin-related lamellae are observed in the highlighted grains at higher B contents [Figure 13(a)]. Figure S10 further quantifies the stacking-fault and twin-boundary atom fractions, normalized by the initial number of γ-phase Ti/Al atoms. During the later stages of deformation, models with higher B contents generally exhibit lower stacking-fault and twin-boundary atom fractions, with a particularly pronounced reduction in the twin-boundary fraction at 1 at.% B. To examine the associated energetics, GSFE profiles were calculated along a two-step twinning path in $\gamma$-TiAl [Figures 13(c-f)]. Increasing the B content from 0 to 1 at.% raises the SISF energy from 0.189 to 0.222 J/m$^2$ and the unstable twinning energy from 0.398 to 0.437 J/m$^2$, while the twinning barrier relative to the SISF state increases only slightly, from 0.209 to 0.214 J/m$^2$. These changes indicate that B increases the energetic cost of the examined faulted configuration and slightly raises the barrier to subsequent twinning, providing an energetic basis for the changes in defect evolution.

Additionally, Figures 13(g-j) show a B-dislocation pinning-depinning event. During this process, the minimum B-dislocation separation decreases from 17.64 to 0.95 Å before increasing to 14.59 Å as the dislocation breaks away. This observation provides direct evidence of local dislocation pinning by B. Overall, the results suggest that B enrichment in disordered regions may increase resistance to grain-boundary-mediated deformation, while modified fault energetics and dislocation pinning provide complementary strengthening mechanisms.

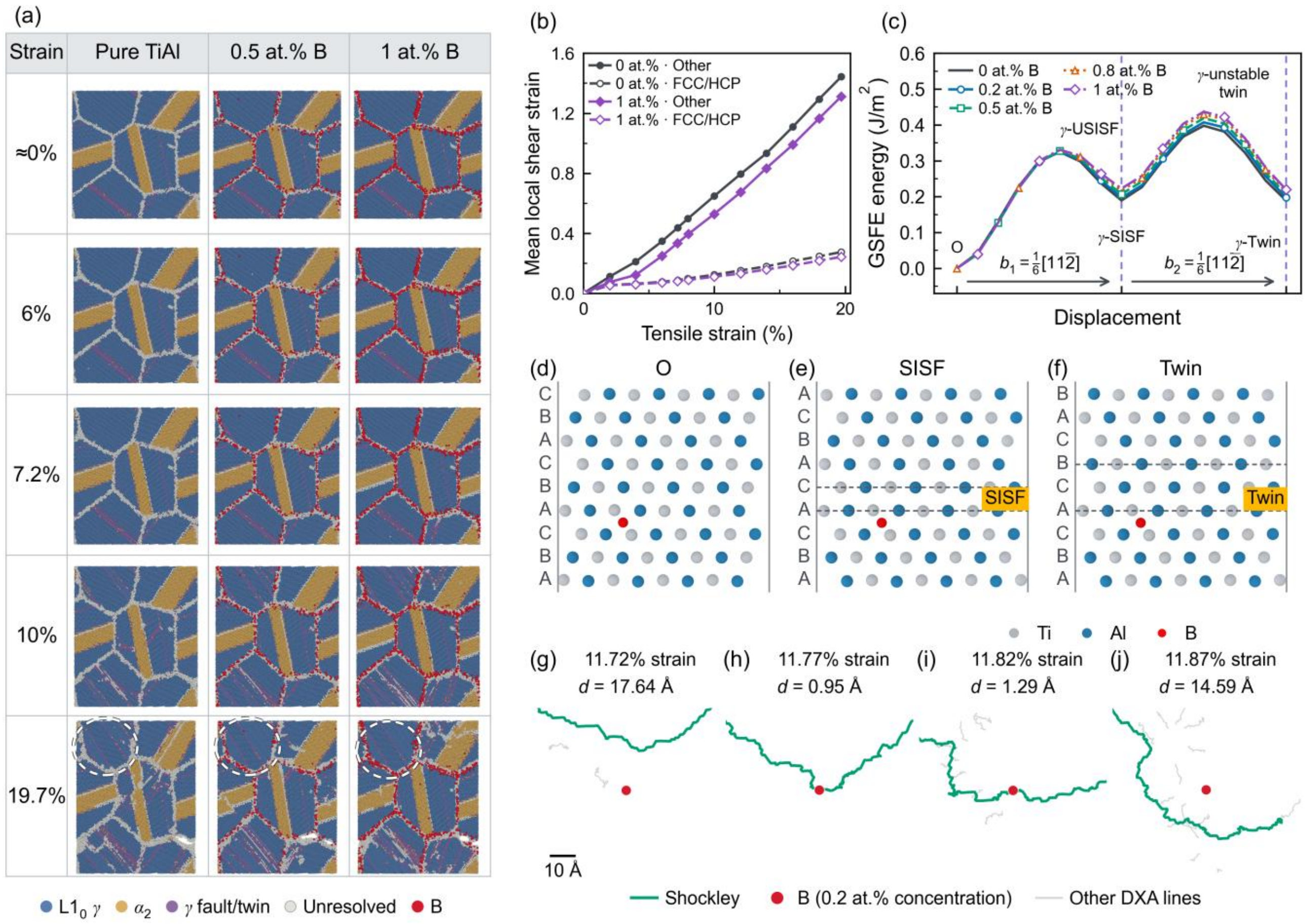

**Figure 13.** (a) Structural evolution at five tensile strains and three B contents at 900 K, using the same fractional-position slice and viewing scale. Colors indicate the $\gamma$ matrix (blue), $\alpha_2$ lamellae (gold), fault/twin-related regions (purple), unresolved regions (gray), and B atoms (red). (b) Mean local shear

strain in PTM Other and FCC/HCP regions for B-free and 1 at.% B TiAl. (c) GSFE profiles along a two-step twinning path at five B contents. (d-f) Original (O), SISF, and twin configurations, with Ti, Al, and B shown in gray, blue, and red. Dashed lines mark faulted or twinned layers. (g-j) Successive configurations of a Shockley partial dislocation (green) interacting with B (red) in the 0.2 at.% B model; d denotes the minimum B-dislocation distance.

## 5. Conclusions

Using an accurate and efficient NEP-TiAlB potential combined with GCMC/MD simulations, we have revealed how local defect environments govern B redistribution and uncovered its mechanical role in TiAl alloys. B preferentially occupies Ti-rich octahedral interstitial sites, where high migration barriers limit its mobility. Al vacancies trap B more strongly than Ti vacancies, while tensile stress and increasing vacancy concentration further promote B enrichment. B preferentially partitions into the $\alpha_2$ phase and exhibits localized enrichment at semi-coherent $\gamma/\alpha_2$ interfaces, whereas no significant segregation is observed at the examined $\gamma/\gamma$ interfaces. B occupation at favorable interfacial sites raises cleavage resistance thereof, linking local chemical environments to interfacial cohesion strength. In the polycrystalline TiAl, B is predominantly enriched in disordered regions. At 900 K, adding 1 at.% B increases the peak stress by 8.65% and the average flow stress by 9.39% relative to B-free TiAl models with identical initial grain and lamellar geometries. B-dislocation interactions and suppressed deformation twinning contribute to this strengthening, which is accompanied by reduced local shear-strain accumulation. Stacking-fault energy calculations further support the observed twinning suppression. These findings demonstrate that B-defect interactions can improve high-temperature deformation resistance at fixed initial grain and lamellar geometries, providing an atomistic basis for tailoring the mechanical performance of TiAl alloys through control of local B occupation and interfacial environments.

## Acknowledgements

This work was supported by the Fundamental and Interdisciplinary Disciplines Breakthrough Plan of the Ministry of Education of China (JYB2025XDXM205), the National Key Research and Development Program of China (2024YFA1409600), the National Natural Science Foundation of China (12225205, U2441272, 1226116036, and 12532020), the Basic Research Program of Jiangsu Province (BK20253025, BK20243044, and BK20253057), the Research Fund of the State Key Laboratory of Mechanics and Control for Mechanical Structures (MCAS-I-0525K01), and the Fundamental Research Funds for the Central Universities. We also acknowledge the support of the High-Performance Computing Center at NUAA. Z.Q.Z. acknowledges support from the Japan Society for the Promotion of Science Postdoctoral Fellowship for Research in Japan (Standard; Fellowship ID: P25373). S.O. and S.S. were supported by the Japan Society for the Promotion of Science (JSPS)

KAKENHI Grant Number JP23K20037 and the Ministry of Education, Culture, Sports, Science and Technology of Japan (MEXT) Programs, Grant Number JPMXP1122684766.

**Author Contributions**

X.M. performed the theoretical calculations and data analysis, and wrote the original manuscript. S.S. provided technical support for the related simulation methods. Z.Q.Z. contributed to the idea generation, research design, data analysis, and writing-review & editing. L.F.Y. contributed to figure layout and design. J.G., L.L.C., and K.O. helped with discussion of the results. S.O. contributed to the simulation methods, data analysis, writing-review & editing. Z.H.Z. contributed to the idea generation, research design, data analysis, writing-review & editing, overall supervision, and funding support. All the authors discussed the results and commented on the manuscript.

**Competing Interests**

The authors declare no competing financial or non-financial interests.

**Data Availability**

The NEP-TiAlB potential will be made publicly available at https://github.com/dream2004/NEP-TiAlB upon publication of this article.

**References**

1. F. Appel, H. Clemens, F. D. Fischer, Modeling concepts for intermetallic titanium aluminides. *Progress in Materials Science* **81**, 55-124 (2016).
2. S. Raji, P. Popoola, S. Pityana, O. Popoola, Characteristic effects of alloying elements on β solidifying titanium aluminides: A review. *Heliyon* **6**, e04463 (2020).
3. P. Janschek, Wrought TiAl Blades. *Materials Today: Proceedings* **2**, S92-S97 (2015).
4. G.-r. Li *et al.*, Progress in Performance Optimization of Ti-Al Alloys for Aviation via Physical Field Application and Alloying. *Journal of Materials Engineering and Performance* **35**, 26141-26162 (2026).
5. D. M. Dimiduk, Gamma titanium aluminide alloys—an assessment within the competition of aerospace structural materials. *Materials Science and Engineering: A* **263**, 281-288 (1999).
6. T. J. Kelly, M. C. Juhas, S. C. Huang, Effect of a B2/gamma structure on the tensile properties of the cast gamma titanium aluminide Ti-48Al-2Cr-2Nb. *Scripta Metallurgica et Materialia* **29**, 1409-1414 (1993).
7. D. Hu, Role of boron in TiAl alloy development: a review. *Rare Metals* **35**, 1-14 (2016).
8. C. Yang, D. Hu, A. Huang, M. Dixon, Solidification and grain refinement in Ti45Al2Mn2Nb1B subjected to fast cooling. *Intermetallics* **32**, 64-71 (2013).
9. M. DeGraef, J. P. A. Löfvander, C. Mccullough, C. G. Levi, The evolution of metastable $B_f$ borides in a Ti-Al-B alloy. *Acta Metallurgica et Materialia* **40**, 3395-3406 (1992).

10. A. B. Godfrey, M. H. Loretto, The nature of complex precipitates associated with the addition of boron to a γ-based titanium aluminide. *Intermetallics* **4**, 47-53 (1996).
11. A. Ahmadian *et al.*, Aluminum depletion induced by co-segregation of carbon and boron in a bcc-iron grain boundary. *Nature Communications* **12**, 6008 (2021).
12. C. T. Liu, C. L. White, J. A. Horton, Effect of boron on grain-boundaries in $Ni_3Al$. *Acta Metallurgica* **33**, 213–229 (1985).
13. J. B. Seol *et al.*, Boron doped ultrastrong and ductile high-entropy alloys. *Acta Materialia* **151**, 366-376 (2018).
14. B. Kang *et al.*, Effect of boron addition on the microstructure and mechanical properties of refractory Al0.1CrNbVMo high-entropy alloy. *International Journal of Refractory Metals and Hard Materials* **100**, 105636 (2021).
15. O. Ouadah, G. Merad, H. S. Abdelkader, Energetic segregation of B, C, N, O at the $\gamma$-TiAl/$\alpha_2$-$Ti_3Al$ interface via DFT approach. *Vacuum* **186**, 110045 (2021).
16. Z.-Z. Li, Y. Wei, H.-B. Zhou, G.-H. Lu, Effects of boron on the mechanical properties of the TiAl–$Ti_3Al$ alloy: A first-principles investigation. *Modern Physics Letters B* **31**, 114593 (2017).
17. G. Kresse, J. Furthmüller, Efficient iterative schemes for ab initio total-energy calculations using a plane-wave basis set. *Physical Review B* **54**, 11169-11186 (1996).
18. J. P. Perdew, K. Burke, M. Ernzerhof, Generalized Gradient Approximation Made Simple. *Physical Review Letters* **77**, 3865-3868 (1996).
19. S. Kirklin *et al.*, The Open Quantum Materials Database (OQMD): Assessing the accuracy of DFT formation energies. *npj Computational Materials* **1**, 15010 (2015).
20. A. Jain *et al.*, Commentary: The Materials Project: A materials genome approach to accelerating materials innovation. *APL Materials* **1**, 011002 (2013).
21. Z. Fan *et al.*, Neuroevolution machine learning potentials: Combining high accuracy and low cost in atomistic simulations and application to heat transport. *Physical Review B* **104**, 104309 (2021).
22. Z. Fan *et al.*, GPUMD: A package for constructing accurate machine-learned potentials and performing highly efficient atomistic simulations. *The Journal of Chemical Physics* **157** (2022).
23. D. Wierstra *et al.*, Natural Evolution Strategies. *Journal of Machine Learning Research* **15**, 949–980 (2014).
24. J. F. Ziegler, J. P. Biersack, "The Stopping and Range of Ions in Matter" in Treatise on Heavy-Ion Science: Volume 6: Astrophysics, Chemistry, and Condensed Matter*,* D. A. Bromley, Ed. (Springer US, Boston, MA, 1985), pp. 93-129.
25. D. Frenkel, B. Smit, "Monte Carlo Simulations in Various Ensembles" in Understanding Molecular Simulation: From Algorithms to Applications. (Academic Press, San Diego, 2002), 10.1016/B978-012267351-1/50007-9 chap. 5, pp. 111-137.
26. P. Hirel, Atomsk: A tool for manipulating and converting atomic data files. *Computer Physics Communications* **197**, 212-219 (2015).
27. J. Ding *et al.*, Micro-mechanism of the effect of grain size and temperature on the mechanical properties of polycrystalline TiAl. *Computational Materials Science* **158**, 76-87 (2019).
28. M. Greenacre *et al.*, Principal component analysis. *Nature Reviews Methods Primers* **2**, 100 (2022).

29. K. Tanaka *et al.*, Elastic constants and their temperature dependence for the intermetallic compound $Ti_3Al$. *Philosophical Magazine A* **73**, 1475-1488 (1996).
30. B. F. Decker, J. S. Kasper, The crystal structure of TiB. *Acta Crystallographica* **7**, 77-80 (1954).
31. S. Otani, Y. Ishizawa, Preparation of $TiB_2$ single crystals by the floating zone method. *Journal of Crystal Growth* **140**, 451-453 (1994).
32. K. E. Spear, P. McDowell, F. McMahon, Experimental Evidence for the Existence of the $Ti_3B_4$ Phase. *Journal of the American Ceramic Society* **69**, C-4-C-5 (1986).
33. P. Nandwana, N. Gupta, S. G. Srinivasan, R. Banerjee, A first principles study of commonly observed planar defects in Ti/TiB system. *Computational Materials Science* **150**, 197-201 (2018).
34. Y. He *et al.*, Elastic constants and thermal expansion of single crystal γ-TiAl from 300 to 750 K. *Materials Science and Engineering: A* **239-240**, 157-163 (1997).
35. Z. Zhao, M. Yi, W. Guo, Z. Zhang, General-purpose neural network potential for Ti-Al-Nb alloys towards large-scale molecular dynamics with ab initio accuracy. *Physical Review B* **110**, 184115 (2024).
36. Y.-K. Kim, H.-K. Kim, W.-S. Jung, B.-J. Lee, Atomistic modeling of the Ti–Al binary system. *Computational Materials Science* **119**, 1-8 (2016).
37. S. Attarian, S. Xiao, Development of a 2NN-MEAM potential for TiB system and studies of the temperature dependence of the nanohardness of $TiB_2$. *Computational Materials Science* **201**, 110875 (2022).
38. O. L. Anderson, A simplified method for calculating the debye temperature from elastic constants. *Journal of Physics and Chemistry of Solids* **24**, 909-917 (1963).
39. Q. Yao, H. Xing, L. J. Meng, J. Sun, Theoretical calculation of elastic properties of $TiB_2$ and TiB. *The Chinese Journal of Nonferrous Metals* **17**, 1297–1301 (2007).
40. K. B. Panda, K. S. R. Chandran, First principles determination of elastic constants and chemical bonding of titanium boride (TiB) on the basis of density functional theory. *Acta Materialia* **54**, 1641-1657 (2006).
41. L. Vitos, A. V. Ruban, H. L. Skriver, J. Kollár, The surface energy of metals. *Surface Science* **411**, 186-202 (1998).
42. Q. Wang *et al.*, First-principles study on the stability and work function of low-index surfaces of $TiB_2$. *Computational Materials Science* **172**, 109356 (2020).
43. Y. Wu, Y. Duan, X. Zhou, M. Peng, Adhesion, stability and electronic properties of TiB (100)/$\alpha$-Ti (0001) coherent interface: Insights from a first-principles investigation. *Surfaces and Interfaces* **30**, 101974 (2022).
44. S. Yang, C. Cui, S. Cui, S. Liu, Y. Liang, Microstructure and mechanical properties of in-situ dual morphology $Ti_8C_5$/$TiB_2$ reinforced TiAl composite. *Materials Science and Engineering: A* **840**, 142918 (2022).
45. J. Han *et al.*, Effect of $TiB_2$ addition on microstructure and fluidity of cast TiAl alloy. *Vacuum* **174**, 109210 (2020).
46. Y. Wang *et al.*, Revealing the phase-interface properties of the $TiB_2$/TiAl composite from a first principles calculations. *Materials Today Communications* **40**, 109730 (2024).

47. W. Li *et al.*, Enhanced mechanical property with refined microstructure of a novel γ-TiAl/TiB2 metal matrix composite (MMC) processed via hot isostatic press. *Materials & Design* **141**, 57-66 (2018).
48. W. Li *et al.*, Enhanced nanohardness and new insights into texture evolution and phase transformation of TiAl/TiB2 in-situ metal matrix composites prepared via selective laser melting. *Acta Materialia* **136**, 90-104 (2017).
49. Y. Yang *et al.*, Effect of scan line spacing on texture, phase and nanohardness of TiAl/TiB2 metal matrix composites fabricated by selective laser melting. *Journal of Alloys and Compounds* **728**, 803-814 (2017).
50. P. A. Burr, E. Kardoulaki, R. Holmes, S. C. Middleburgh, Defect evolution in burnable absorber candidate material: Uranium diboride, UB2. *Journal of Nuclear Materials* **513**, 45-55 (2019).
51. G. Henkelman, B. P. Uberuaga, H. Jónsson, A climbing image nudged elastic band method for finding saddle points and minimum energy paths. *The Journal of Chemical Physics* **113**, 9901-9904 (2000).
52. S. V. Divinski, F. Hisker, T. Wilger, M. Friesel, C. Herzig, Tracer diffusion of boron in α-Ti and γ-TiAl. *Intermetallics* **16**, 148-155 (2008).
53. X. He *et al.*, Unravelling the interplay between hydrogen and grain boundary of *α*-Fe under different concentration and strain rate via neural network interatomic potential. *Computational Materials Science* **267**, 114593 (2026).
54. T. D. Doležal, R. Freitas, J. Li, Complementary thermodynamic mechanisms of boron and carbon segregation at grain boundaries in nickel alloys. *npj Computational Materials* **12** (2026).
55. M. Li, S. Xiao, Y. Chen, L. Xu, J. Tian, The effect of boron addition on the high-temperature properties and microstructure evolution of high Nb containing TiAl alloys. *Materials Science and Engineering: A* **733**, 190-198 (2018).